\documentclass[prc, 12pt, preprint,nofootinbib]{revtex4}
\usepackage{color}
\usepackage{graphicx}
\usepackage{array,longtable}

\begin{document}

\title{Physics of even-even superheavy nuclei with 96 $<$ Z $<$ 110 in the Quark-Meson-Coupling Model.}
\author{J. R. Stone~\footnote{jirina.stone@physics.ox.ac.uk}}
\affiliation{ Department of Physics (Astro), University of Oxford, Keble Road OX1 3RH, Oxford, United Kingdom, \\
 Department of Physics and Astronomy, University of Tennessee, Knoxville TN 37996, USA} 
\author{K. Morita~\footnote{moritako@phys.kyushu-u.ac.jp}}
\affiliation { Department of Physics, Kyushu University, Nishi-ku, Fukuoka 819-0395, Japan, \\
RIKEN Nishina Center, RIKEN, Wako-shi, Saitama 351-0198, Japan}
\author{P. A. M. Guichon~\footnote{pierre.guichon@cea.fr}}
\affiliation{ CEA/IRFU/SPhN Saclay, F91191 France}
\author{A. W. Thomas~\footnote{anthony.thomas@adelaide.edu.au}}
\affiliation { CSSM and CoEPP, Department of Physics, University of Adelaide, SA 5005 Australia}
\date{\today}
\begin{abstract}
The Quark-Meson-Coupling (QMC) model has been applied to the study of the properties of even-even superheavy nuclei with 96 $\leq$ Z $ \leq$ 110, over a wide range of neutron numbers. The aim is to identify the deformed shell gaps at N=152 and N=162 predicted in macroscopic-microscopic (macro-micro) models, in a model based on the mean-field Hartree-Fock+BCS approximation. 

The predictive power of the model has been tested on proton and neutron spherical shell gaps in light doubly closed (sub)shell nuclei $^{40}$Ca, $^{48}$Ca, $^{56}$Ni,  $^{56}$Ni , $^{78}$Ni , $^{90}$Zr, $^{100}$Sn , $^{132}$Sn, $^{146}$Gd  and $^{208}$Pb, with results in a full agreement with experiment.

In the superheavy region, the ground state binding energies of 98 $\leq$ Z $\leq$ 110  and 146 $\leq$ N $\leq$ 160 differ, in the majority of cases, from the measured values by less than $\pm$2.5 MeV, with the deviation decreasing with increasing Z and N. The axial quadrupole deformation parameter, $\beta_{\rm 2}$, calculated over the range of neutron numbers 138 $\leq$ N $\leq$ 184, revealed a prolate-oblate coexistence and shape transition around N=168, followed by an oblate-spherical transition towards the expected N=184 shell closure in Cm, Cf, Fm and No. The closure is not predicted in Rf, Sg, Hs and Ds as another shape transition to a highly deformed ($\beta_{\rm 2} \sim$ 0.4) shape in Sg, Hs and Ds for N$>$ 178 appears, while $^{288}$Rf (N=184) remains oblate.

The bulk properties predicted by QMC, such as ground state binding energy, two-neutron separation energy, the empirical shell-gap parameter  $\delta_{\rm 2n}$ and Q$_\alpha$ values, are found to have a limited sensitivity to the deformed shell gaps at N=152 and 162. However, the evolution of the neutron single-particle spectra with  0 $\leq \beta_{\rm 2} \leq$ 0.55 of  $^{244}$Cm,  $^{248}$Cf, $^{252}$Fm, $^{256}$No, $^{260}$Rf, $^{264}$Sg, $^{268}$Hs and $^{272}$Ds, as representative examples, gives a (model dependent) evidence for the location and size of the N=152 and 162 gaps as a function of Z and N. In addition, the neutron number dependence of neutron pairing energies provides supporting indication for existence of the energy gaps.

Based on these results, the mean-field QMC and macro-micro models and their predictions of deformed shell structure of superheavy nuclei are compared. Clearly the QMC model does not give results as close to the experiment as the macro-micro models. However, considering that it has only  four global variable parameters (plus two parameters of the pairing potential), with no local adjustments, the results are promising.
\end{abstract}
\maketitle

\section{Introduction }
\label{intro}
The stability of super-heavy elements (SHE) and their isotopes is critically dependent on their shell structure. Location of shell gaps is essential in the search for new elements and to understand better their isotopes and isomers. Beyond the classical proton and neutron spherical shell gaps at Z=82 and N=126, nuclei are increasingly less stable until  the gaps predicted at Z=114 and N=178--184 are reached. The majority of nuclei in the transuranium  region are deformed and thus theoretical predictions of location of stabilizing deformed shell gaps becomes important. This task is challenging because the level density in deformed heavy nuclei increases as compared to lighter nuclei and shell gaps are typically reduced to hundreds of keV. The location and size  of the gaps is model dependent and extremely demanding on the accuracy of calculation procedures.

In the region of interest of this paper, deformed shell gaps near Z=100 and N=152 and  Z=108 and N=162 have been predicted in macro-micro models, see e.g. \cite{patyk1991,moller1994,cwiok1994,smolanczuk1995,muntian2001,muntian2003,muntian2003a,baran2005,jachimowicz2014,moller2016}, as well as a preformed $\alpha$ particle model with M3Y interaction \cite{ismail2010}.  Mean-field energy density functionals based on Skyrme and Gogny forces (see e.g. \cite{cwiok1996,bender2002,bender2003,ward2013,shi2014,shi2014a,heenen2015,dobaczewski2015,ward2015} ) and relativistic mean field theory \cite{greiner2002,afanasjev2003,afanasjev2013,afanasjev2016} provided a variety of neutron shell gaps but have not converged to the same results as macro-micro models, except for \cite{shi2014}, with some local adjustment of the UNEDF type interactions. 

Experimental evidence for increased stability related to shell gaps is mainly based on $\alpha$-decay life-times and energies of emitted $\alpha$-particles, as well as data on spontaneous fission.  Lazarev {\em et al.} first reported enhanced stability near the deformed shells N=162 and Z=108 from studies of $^{265,266}$Sg (Z=106, N=159, 160) \cite{lazarev1994}, $^{167}$Hs (Z=108, N=159) \cite{lazarev1995} and decay products of $^{273}$Ds (Z=110, N=163) \cite{lazarev1996}. A topical review of both theoretical models and experimental data available before 2007 in the regions of Z $\leq$ 113 and 112 $\leq$ Z $\leq$ 118 emphasized  the general concept that enhanced stability and shell gaps are related \cite{oganessian2007} . Survey of $\alpha$-decay energies from Fm (Z=100), No (Z=102) and Rf (Z=104) showed an increase in $\alpha$-decay lifetime at N=152 and in Sg, Hs and Ds at N=162.  Oganessian {\em et al.} \cite{oganessian2013} observed enhancement of the partial spontaneous fission half-life in  $^{254}$No (N=152) and elements with 102 $<$ Z $<$ 108 with N=162. Hofmann {\em et al.} identified a new isotope, $^{270}$Ds (N=160), and its decay products  $^{266}$Hs (N=158) and  $^{262}$Sg (N=156) \cite{hofmann2001}, giving more evidence for increased stability in this region. Nishio {\em et al.} discovered a new isotope, $^{268}$Hs (N=160) \cite{nishio2010}, and obtained  the Q$_\alpha$ value for its decay to $^{264}$Sg (N=160).  Dvorak {\em et al.} \cite{dvorak2006} observed a new nuclide $^{270}$Hs (N=162) which decays by  $\alpha$-particle emission to $^{266}$Sg. $^{270}$Hs has become the first nucleus for which experimental nuclear decay properties are available for comparison with theoretical predictions of the N=162 shell stability.  $^{270}$Hs was later revisited \cite{oganessian2013a} and a Q$_\alpha$ value, consistent with a shell gap at Z=108 and N=162, was obtained. 

In addition to data on $\alpha$-decays and spontaneous fission, ground state binding energies and their differences serve  as a source of information on global trends of nuclear shell structure. Until very recently most of the binding energies were obtained from decay spectroscopy. This method requires at least one absolute mass to be known in a decay chain, to which one can relate the masses of decay products. This works best for even-even nuclei, where the decays are likely to proceed between the parent and daughter ground states. It is more problematic for decay chains involving odd-A and odd-odd nuclei, in which decays from and to excited states may be involved and $\beta$-decay branching is possible. Also, in longer decay chains the errors inevitably accumulate, decreasing the accuracy of masses deduced at the bottom of the chain. Therefore an independent direct measurement of atomic masses yielding related binding energies is very desirable. Relevant to this work, recent direct mass measurement of $^{252-255}$No and $^{255,256}$Lr with SHIPTRAP provided solid evidence for a manifestation of the shell gap at N=152 in No and Lr isotopes \cite{block2015}. 

In the present work we use the latest comprehensive compilation of binding energies (mainly obtained indirectly from spectroscopy or reactions) and related quantities presented in AME2016 \cite{wang2017}, as experimental data wherever applicable.

 We employ the Quark-Meson-Coupling energy density functional (QMC EDF)  QMC-$\pi$-I to examine detailed ground state properties of even-even spherical and axially symmetric deformed nuclei with 96 $\leq$ Z $\leq$ 110 and 138 $\leq$ N $\leq$ 184.  This work is a continuation of the more global application of the same version of the  QMC model in the region 96 $\leq$ Z $\leq $ 136 and 118 $\leq$ N $\leq$ 320 \cite{stone2017}. The model is an extension of the QMC-I model, applied to finite nuclei below $^{208}$Pb \cite{stone2016}, by inclusion of the pion exchange Fock term. The basic idea of the model is that, instead of the traditional modeling of nuclear forces through exchange of mesons coupled to point-like nucleons, this exchange takes place directly between quarks in different nucleons. The nucleons have internal structure, taken as a cluster of valence quarks in a confining potential such as, but not limited to, the MIT bag \cite{degrand1975}. When the quarks in one nucleon interact self-consistently with the quarks in the surrounding nucleons by exchanging a $\sigma$ meson, the effective mass M$_{\rm N}^\star$ of the nucleon is no longer linear in the scalar mean field $<\sigma>$. It is expressed as
\begin{equation}
M_{\rm N}^\star = M_{\rm N} - g_{\rm \sigma N} \sigma + (d/2) (g_{\rm \sigma N}\sigma)^2,
\end{equation}
where g$_{\rm \sigma N}$, the $\sigma$-nucleon coupling constant in free space, is a parameter of the model. By analogy with electromagnetic polarizabilities, the coefficient $d$, calculated in terms of the nucleon internal structure, is known as the “scalar polarizability” \cite{guichon1988,guichon1996}. The appearance of this term in the nucleon effective mass is sufficient to lead to the saturation of nuclear binding thus demonstrating a link between the internal structure of the nucleon and fundamental properties of atomic nuclei.

The results presented here include binding energies E$_{\rm b}$, deformation parameters $\beta_{\rm 2}$, two-neutron separation energies S$_{\rm 2n}$,  Q$_\alpha$ values, empirical shell gap parameter $\delta_{\rm 2n}$ and single-particle spectra, calculated in the self-consistent, mean-field Hartree-Fock (HF) approximation with a BCS pairing model. A full description of the model can be found in \cite{guichon2018} and references therein.

We keep to the minimum a comparison of QMC results with the outcome of other theoretical models. We believe that the only prove of correctness of a theory is agreement with reliable experimental data. There are too many theoretical models in the literature and too few reasons for choosing one over the other. It is beyond the scope of this work to tackle this task and we leave the reader to the quoted references. As an exception, we employ the macro-micro models, the macroscopic Finite-Range-Droplet (FRDM) with the folded-Yukawa single-particle microscopic part \cite{moller2016}  and the MM model, which uses the Yukawa+exponential model for the macroscopic part and the deformed Woods-Saxon potential to calculate the shell correction energy \cite{muntian2001,muntian2003,muntian2003a}, for a comparison. These models are widely used in the SHE community and provide a benchmark for a comparison with the QMC calculations. Interesting discussion of performance of the two models can be found, for example, in \cite{hofmann2016}. A short comment on single-particle spectra calculated in selected Skyrme and RMF models is made in Sec.~\ref{heavy}.

\section{Bulk properties}
\label{bulk}
\subsection{Calculation method}
\label{calmeth}
In the self-consistent HF method, an empirical EDF of a system of N particles, describing the total energy density of the system in terms of adjustable parameters, is employed. Using the variational principle, the N-body problem is reduced to a set of N single-particle HF differential equations. This set is solved iteratively to yield single-particle eigen-energies and eigen-functions describing the ground state of the system, subsequently used to calculate particle densities and observables such as binding energy, charge and mass density distribution and their moments. The minimization process starts from selected wavefunctions, derived from a simple potential such a folded Yukawa, Woods-Saxon or harmonic oscillator, and proceeds either without any additional constraint of the path to the final minimum \cite{vautherin1973}, or with an applied a constraint (CHF) usually requiring, but not limited to, a fixed value of the quadrupole moment \cite{flocard1973}. In practice, for axially deformed nuclei, the latter procedure involves determination of the equilibrium wavefunctions and single-particle energies at each chosen value of the deformation parameter $\beta_{\rm 2}$ used to calculate the quadrupole moment $<Q_{\rm 2}>$=(3/4$\pi$)AR$^{\rm 2}_{\rm 0}\beta_{\rm 2}$, with A being the mass number and R$_{\rm 0}$ = 1.2 fm. Changing the deformation parameter by a fixed amount through an expected range of deformations yields the lowest energy of the system and its equilibrium shape.  

The binding energies presented in this work have been obtained using the CHF method. For each isotope, $\beta_{\rm 2}$ was varied on a mesh with a step of 0.02 from -0.5 to 0.6 and the binding energy E$_{\rm_b}$ calculated at each point. This procedure introduced an estimated average uncertainty of the order of tens of keV or less in the binding energy and $\sim$0.02 in $\beta_{\rm 2}$ . In the attempt to reduce these uncertainties, a five-point fit to a parabola was performed  taking the mesh equilibrium value and two neighbouring points each side. We found the correction from the fit was small, as demonstrated in Sec.~\ref{search_bulk} and, unless stated otherwise, results from the mesh calculations were adopted. 

 The HF+ BCS code SKYAX allowing for axially symmetric and reflection-asymmetric shapes, adapted by P.-G. Reinhard \cite{reinhard,stone2016} for the QMC EDF, was used. The BCS approximation requires {\em a priory} knowledge of the pairing potential and is applicable only in the case of time-reversal invariance, i.e. for even-even nuclei. In this case the time-conjugate energies and wave functions of a nucleon are degenerate, $e_\alpha = e_{\bar\alpha}$ and $\phi_\alpha = \phi_{\bar \alpha}$. The pairing potential is forced to be diagonal in the basis $\phi_\alpha$ leading to a two-component wavefunctions $u_\alpha\phi_\alpha$ and $v_\alpha\phi_\alpha$ with the single quasiparticle eigen-energy $\epsilon_\alpha$.  The pairing problem is thus reduced to determination of occupation amplitudes $u_\alpha$ and $v_\alpha$ by solution of the gap equation \cite{bender2003}. In this work we term the energies $\epsilon_\alpha$ ``single-particle states''.

At sphericity, the single-particle states, in the harmonic oscillator approximation, are labeled by quantum numbers [$n \ell j$] with $j=\ell \pm s$. 
For axially deformed shapes, the anisotropic harmonic oscillator model is adopted. It is convenient to use cylindrical coordinates ${r,z}$. Taking the $z$-axis to be the symmetry axis, the z-component of the orbital momentum  $m_\ell$, spin $m_s$ and of the total angular momentum $j_z$ (with eigenvalues $\Omega=m_\ell + m_s = m_\ell \pm 1/2$)  are good quantum numbers. The main oscillator number in these coordinates is $N=n_z+2n_r+m_\ell$. The single-particle states are labelled $\Omega ^\pi [N n_z m_\ell]$ where the parity is $\pi=(-1)^\ell=(-1)^N$ \cite{ringandschuck}. In this work the single-particle states were calculated in an equivalent basis $ [N\ell\Lambda \Omega]$  with $\Lambda$ and $\Sigma$ being projections of the orbital and spin momenta to the z-axis \cite{Bohr}. In this basis each state has two components, one with $\Omega=\Lambda+1/2$ and one with $\Omega=\Lambda-1/2$. For example, for $\ell=2$ and $\Omega=3/2$  the wave function has two components, $d_{5/2,3/2}$ and $d_{3/2,3/2}$. There is only one component for $\Omega=5/2$, $d_{5/2,5/2}$, as $d_{3/2,5/2}$ is not allowed. The single-particle states are labeled as $\Omega^\pi [N\ell\Lambda \Omega]$. 

The best parameter set was sought using the experimental data set chosen by Klüpfel {\em et al.} \cite{klupfel2009}. The proton and neutron pairing strengths  of the pairing force in the BCS model were also fitted to data in \cite{klupfel2009}. We note that the addition of the explicit pion exchange in the model did not increase the number of variable parameters beyond the four used in \cite{stone2016}. The coupling strengths G$_\sigma$ , G$_\omega$ , G$_\rho$ were obtained from a fit to experiment to be 11.15, 8.00 and 7.38  fm$^{\rm 2}$, respectively, and the mass of the $\sigma$ meson M$_\sigma$=712 MeV.  In addition, the g$_{\rm A}$=1.26 and the effective pion constant $f_\pi$=0.471 were kept fixed. The new parameter set is compatible with nuclear matter properties, the saturation energy  E$_{\rm 0}$=-15.8 MeV,  the saturation density $\rho_{\rm 0}$=0.153 fm$^{\rm −3}$, incompressibility K$_{\rm 0}$=319 MeV, and the symmetry energy and its slope at saturation S$_{\rm 0}$=30 MeV and L=27 MeV.

The standard pairing energy functional in the BCS model with $\delta$ function pairing interaction, acting through the whole nuclear volume, has the form \cite{stone2007}
 \begin{equation}
E_{\rm pair} = \frac{1}{4}\sum\limits_{q\in(p,n)}{V^{\rm pair}_q \int{d^3r\chi^2_q}}, \,\,\,\,\,\, \chi_{\rm q}(\vec{r})=
\sum\limits_{\alpha\in{q}} {w{_\alpha}u{_\alpha}v{_\alpha}|\phi_{\alpha}(\vec{r})|^2}
\end{equation}
where q$\in$(p,n) and the pairing strengths $V^{pair}_{q}$  have to be fitted to experimental data, thus increasing the number of the QMC model parameters when applied to finite nuclei. The functions $\chi_{\rm q}$ depend on a phase-space factor $w_\alpha$ \cite{klupfel2009}, occupation probabilities $u_\alpha$ = $\sqrt{(1-v_\alpha^2}$ and $v_\alpha \in  (0,1)$ and the single-nucleon states $\phi_\alpha$. The summation is performed over all occupied states.

\subsection{Ground state binding energies and quadrupole deformations}
\label{binding}
\subsubsection{Binding energies}
\label{be}
Ground state atomic masses and their differences play a major role in predictions of transmutation probabilities of isotopes of superheavy elements. In this work we calculate total ground state binding energies, $E^{cal}_{b}$(Z,N) which should be compared to the experimental values, $E^{exp}_{b}$(Z,N), obtained from the atomic mass excess M(Z,N) as \cite{moller1997}
\begin{equation}
E_b=Z M(^1H)+NM(^1n)-M(A,Z)
\end{equation}
where M$(^1H)$ is the hydrogen-atom mass excess 7.289034 MeV and $M(^1n)$ is the neutron mass excess 8.071431 MeV. 

Macro-micro models such  as  MM \cite{muntian2001,muntian2003,muntian2003a} and FRDM \cite{moller2016} are known to predict binding energies closer to experiment than conventional mean-field models with effective density dependent interactions in the HF+BCS approximation in the SHE region.

We summarize in Table~\ref{tab:1} the difference between the experimental binding energy $E^{exp}_b $, as listed in AME2016 \cite{wang2017}, and values of the QMC, FRDM and MM models in the region of 98 $\leq$ Z $\leq$ 110 and 146 $\leq$ N $\leq$ 160. The majority of the QMC predictions deviate from experiment by less than $\pm$2.5 MeV, except for $^{\rm 252,254}$Cf. The FRDM yield binding energies deviating by within less than 1 MeV from experiment, except for $^{\rm 266}$Hs and $^{\rm 270}$Ds. Both models exhibit a slight overbinding with increasing Z and N.  The available MM data are limited to 102 $\leq$ Z $\leq$ 110 and the largest difference between experiment and theory is 340 keV.  We note that the QMC results have been obtained {\em without any local adjustment} specific to the SHE region and the SHE data were not included in the fitting of QMC four parameters. The values of ten constants of the FRDM were determined directly from an optimization to fit ground-state masses of 2149 nuclei which included known superheavy masses up to $^{265}$Sg and $^{264}$Hs \cite{moller2016} and the MM model was specially adapted for heavy nuclei \cite{muntian2001}. It should also be mentioned that the difference between experiment and theory in the QMC model decreases with increasing Z and N which is the opposite trend to that of the FRDM model.  

\subsubsection{Quadrupole deformation}
\label{gd}
Knowledge of SHE isotope shapes is important because the probability of their decay depends not only on the differences in binding energies but also on the structure of parent and daughter isotopes, determining their shape. 

 Experimental evidence on the energies of the first excited 2$^+$ and 4$^+$ states, which carry information on the ground state deformation, is available only in $^{240}$U (Z=92), \cite{ishii2005}, $^{244}$Pu (Z=94) \cite{IAEA}, $^{248}$Cm (Z=96) \cite{czosnyka1986}, $^{252}$Cf (Z=98) (\cite{takahashi2010} and refs. therein) and  $^{256}$Fm \cite{IAEA}, but no data exist for heavier elements. The energies are remarkably similar and their ratios close to the value of 3.3, expected for an axially symmetric rigid rotor. The heaviest nuclei with known sign and magnitude of their ground state quadrupole moments are  $^{233,235}$U, $^{241}$Pu,  $^{241,243}$Am (Z=95) and $^{253}$Es (Z=95) \cite{nstone2016}. They are all consistent with a large prolate deformation.

\begin{figure}
\includegraphics[width=18cm]{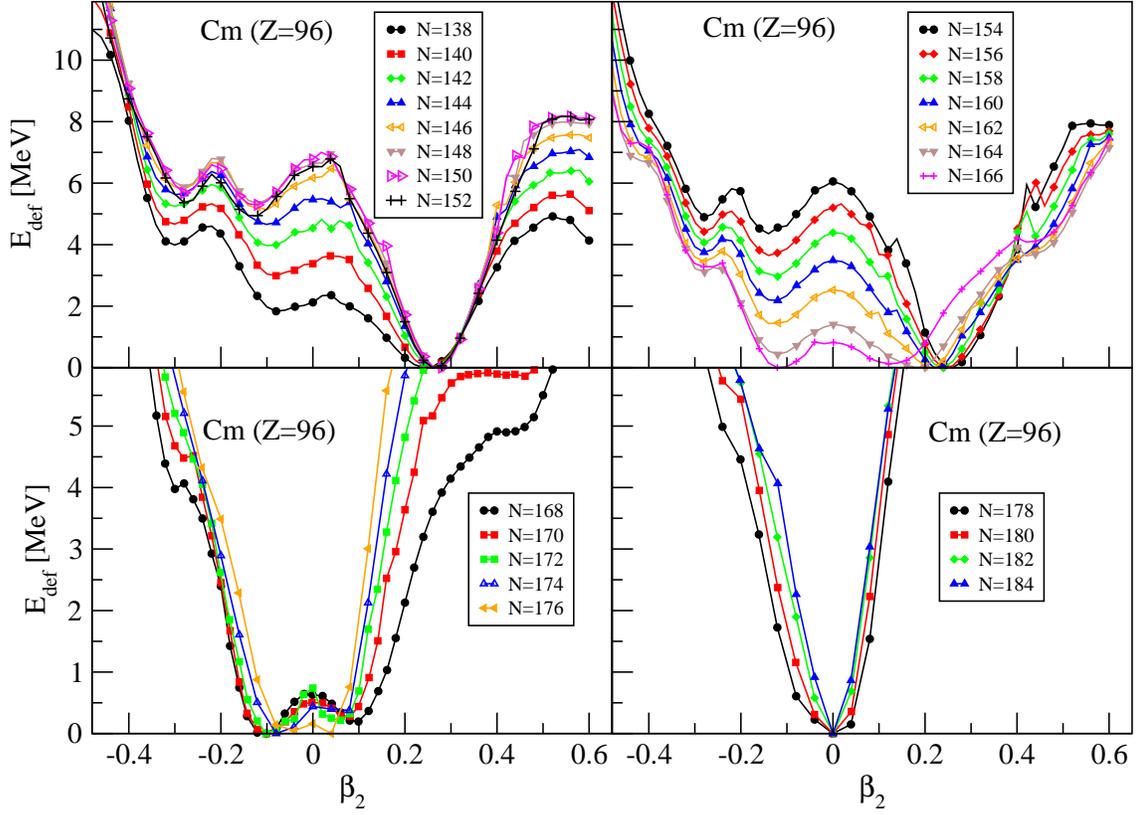}
\centering{}\caption{Deformation energy $E_{def}$ as a function of the quadrupole deformation parameter $\beta_{\rm 2}$ as calculated for $^{\rm 244}$Cm in the QMC model. The colour code is independent in each panel. Note that the y-axis scale in the bottom panels is a double the scale in the top panels. For more information see the text.}
\label {fig:1}
\end{figure}

\begin{figure}
\includegraphics[width=18cm]{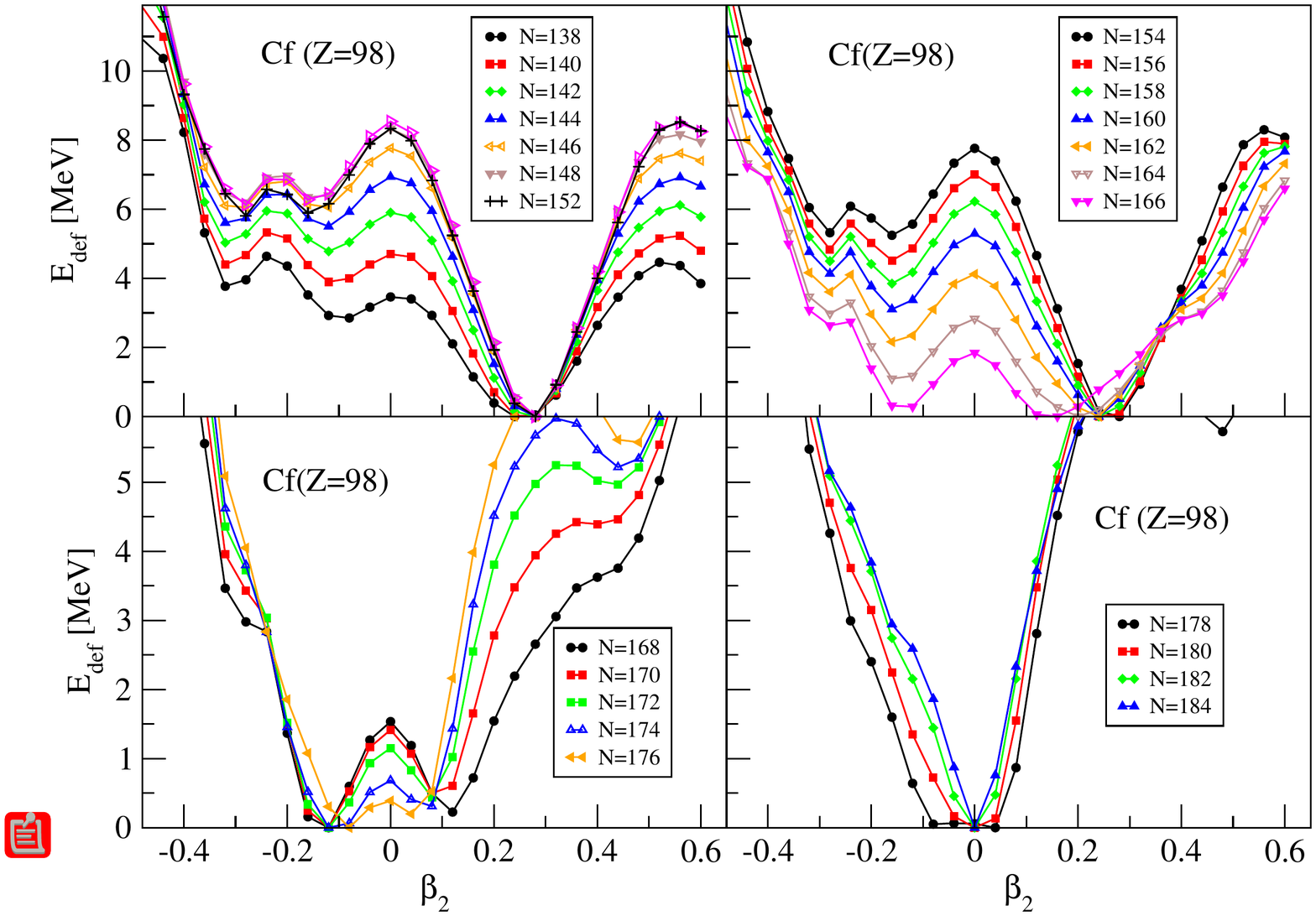}
\centering{}\caption{The same as Fig.\ref{fig:1} but for $^{\rm 248}$Cf.}
\label {fig:2}
\end{figure}

\begin{figure}
\includegraphics[width=18cm]{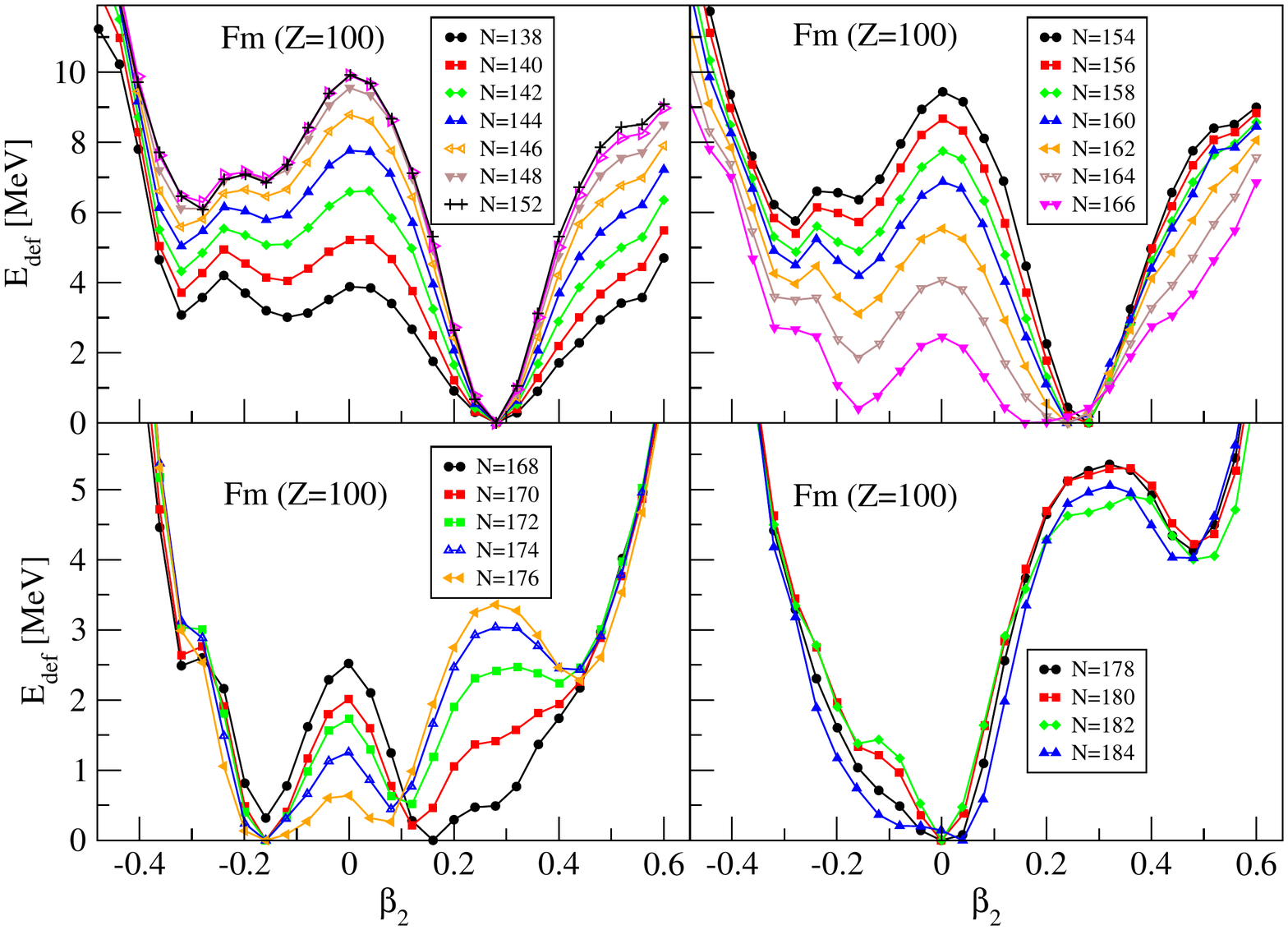}
\centering{}\caption{The same as Fig.\ref{fig:1} but for $^{\rm 252}$Fm.}
\label {fig:3}
\end{figure}

\begin{figure}
\includegraphics[width=18cm]{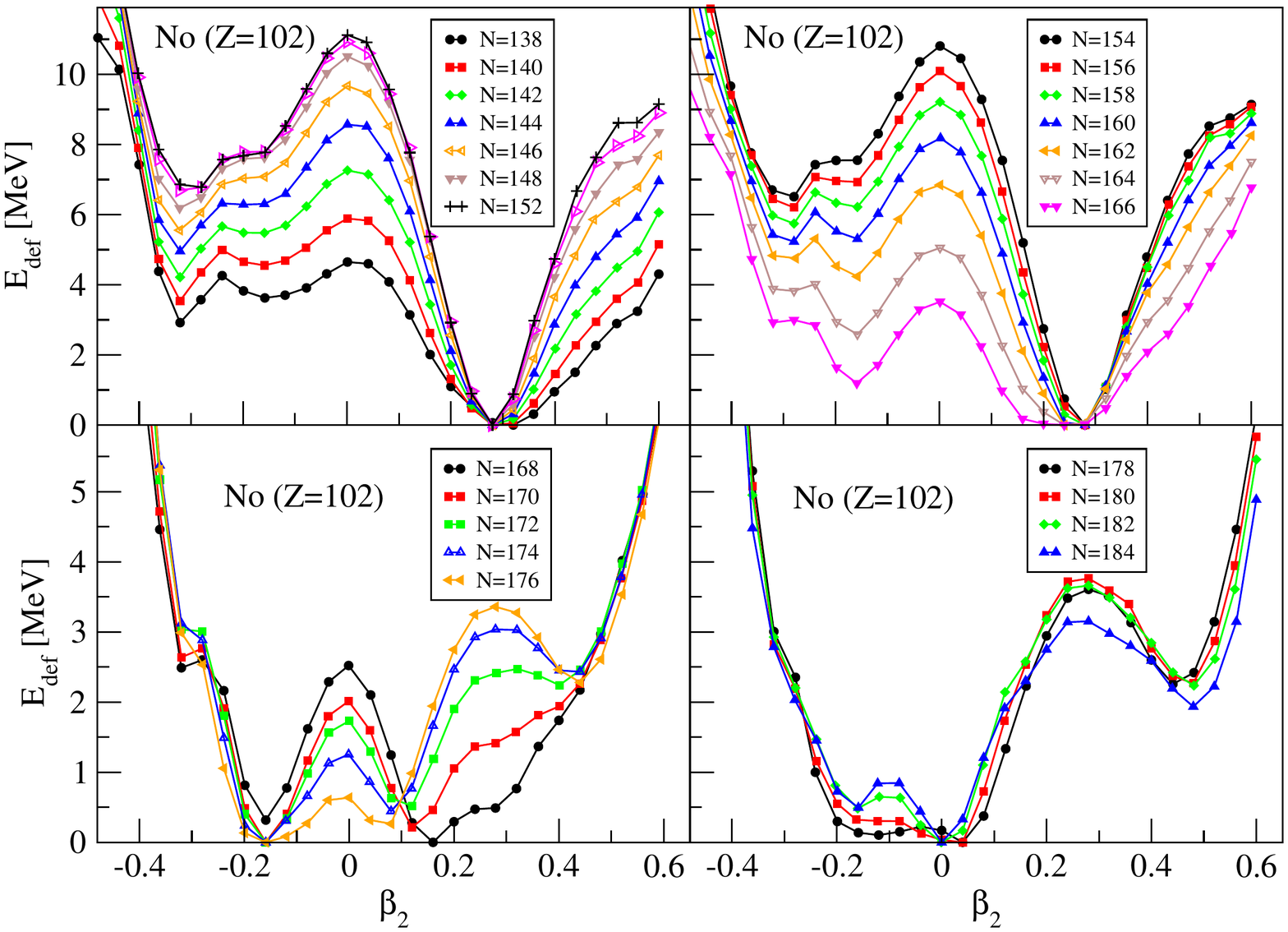}
\centering{}\caption{The same as Fig.\ref{fig:1} but for $^{\rm 256}$No.}
\label {fig:4}
\end{figure}

\begin{figure}
\includegraphics[width=18cm]{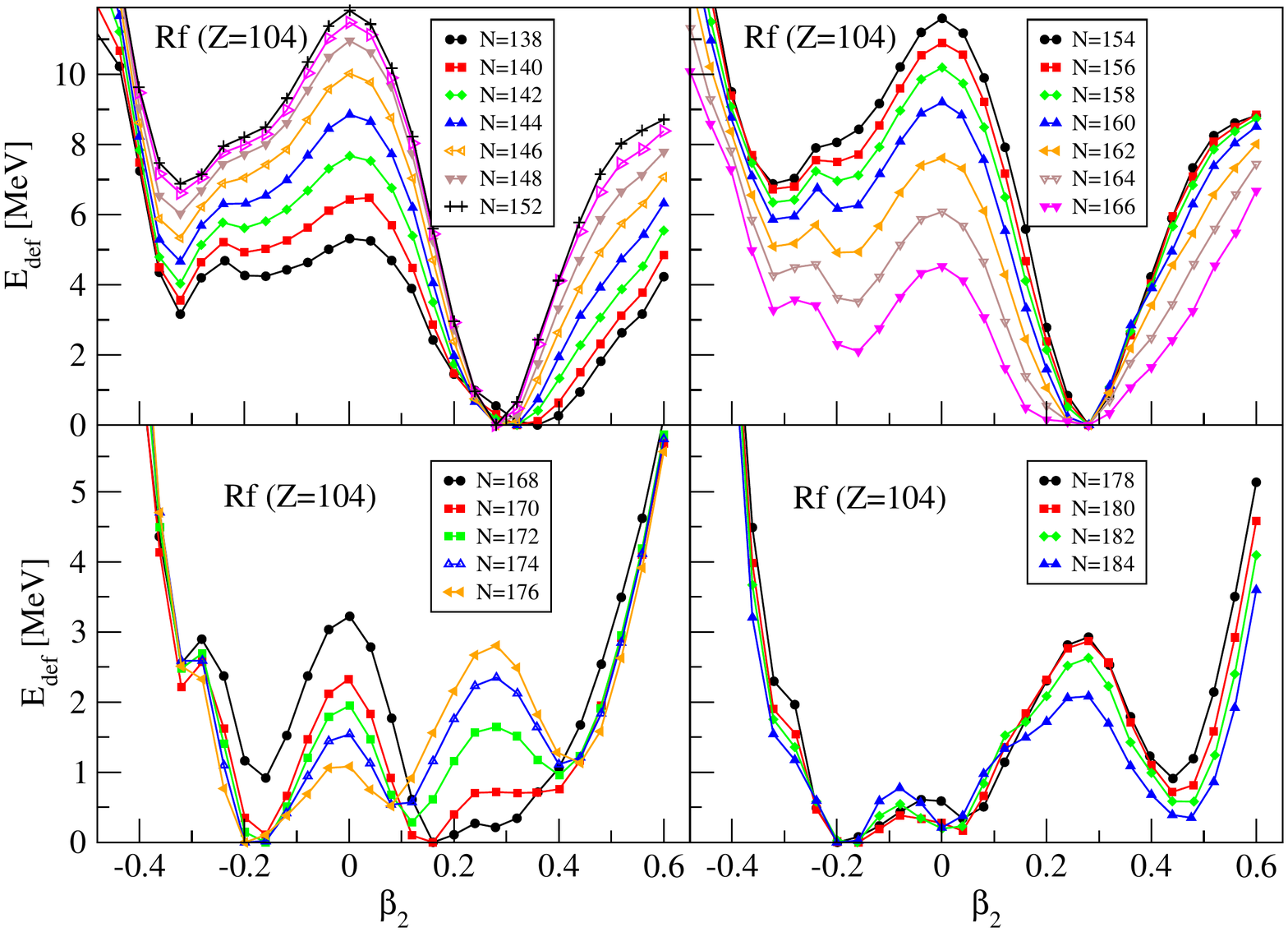}
\centering{}\caption{The same as Fig.\ref{fig:1} but for $^{260}$Rf.}
\label {fig:5}
\end{figure}

\begin{figure}
\includegraphics[width=18cm]{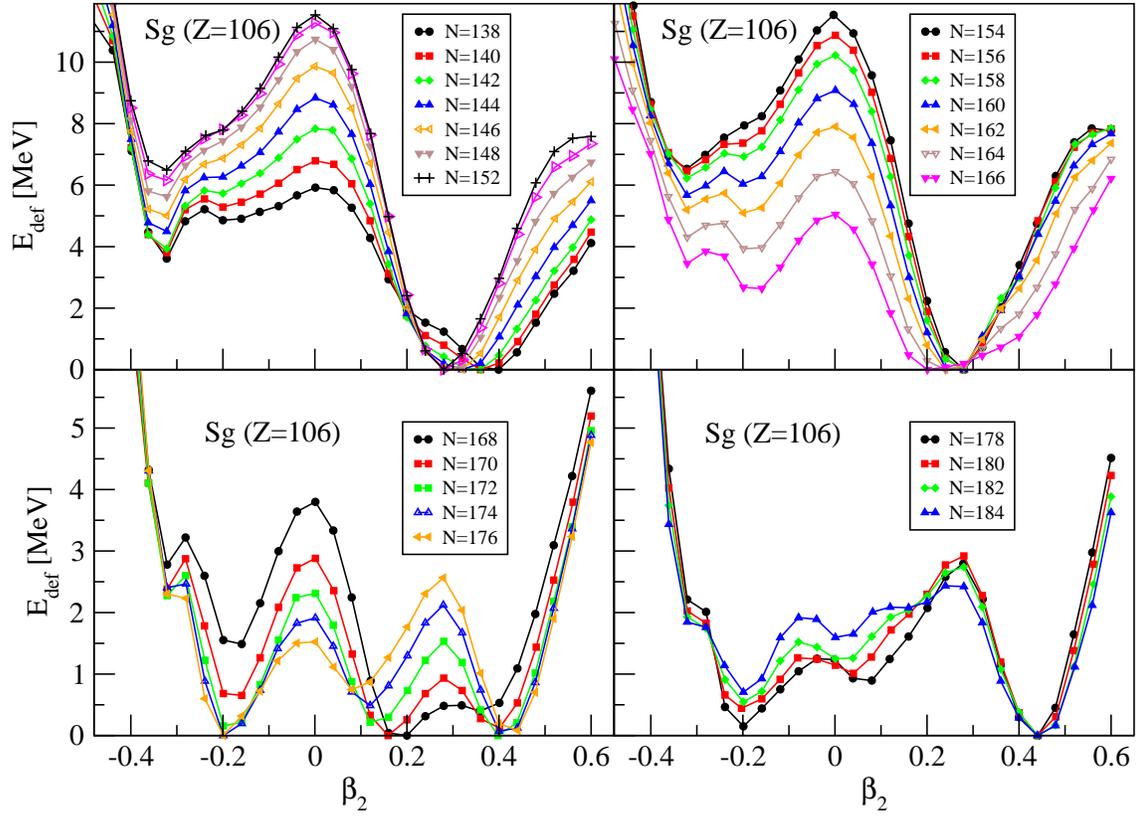}
\centering{}\caption{The same as Fig.\ref{fig:1} but for $^{\rm 264}$Sg.}
\label {fig:6}
\end{figure}

\begin{figure}
\includegraphics[width=18cm]{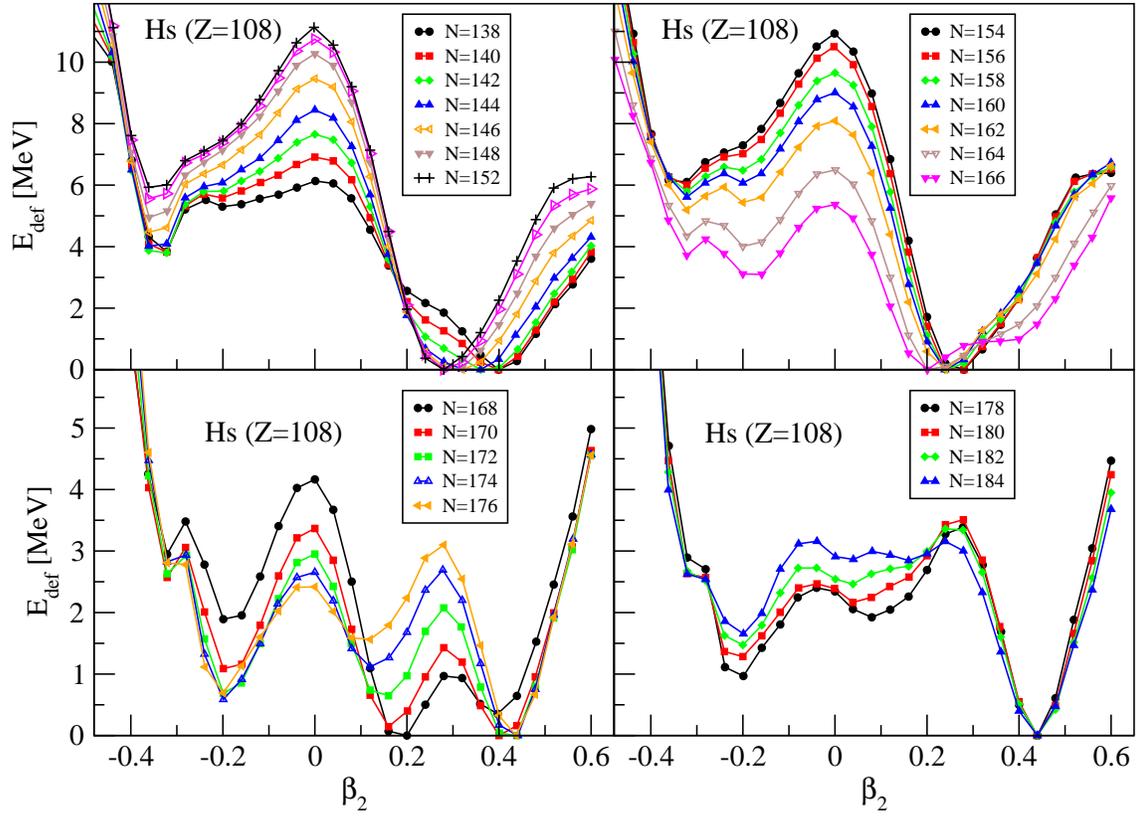}
\centering{}\caption{The same as Fig.\ref{fig:1} but for $^{\rm 268}$Hs.}
\label {fig:7}
\end{figure}

\begin{figure}
\includegraphics[width=18cm]{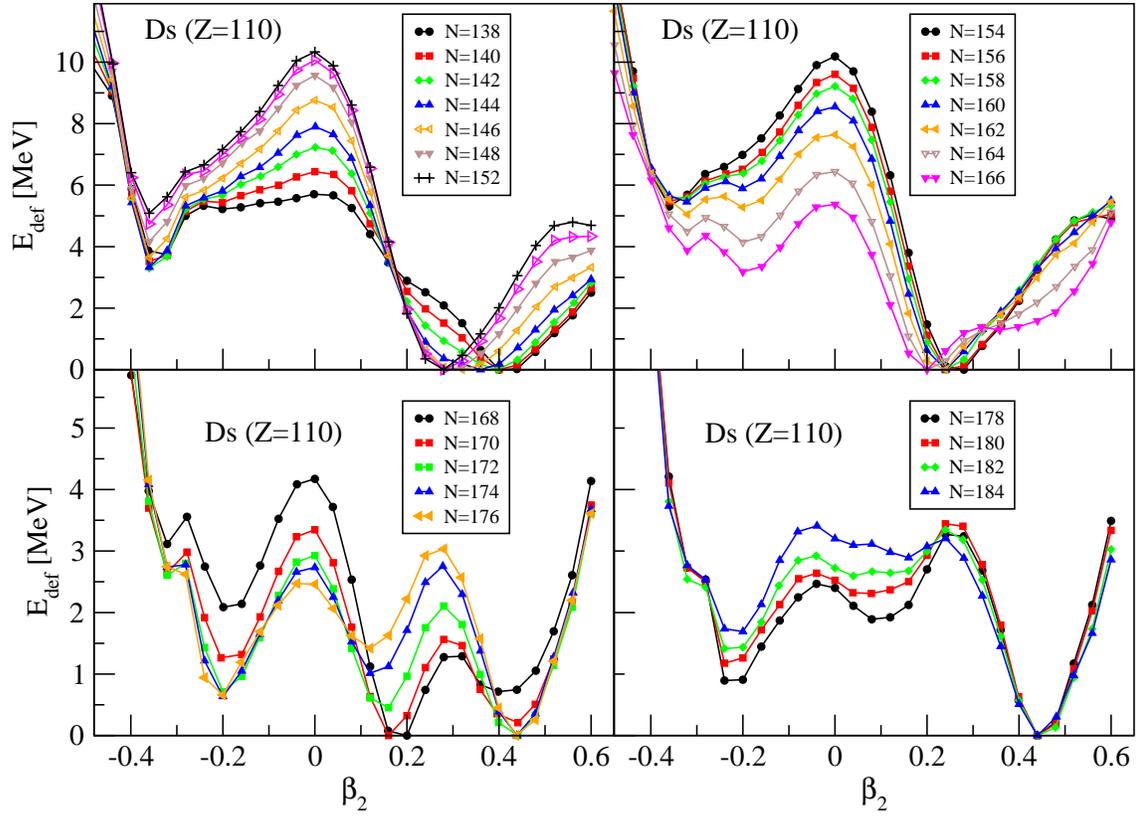}
\centering{}\caption{The same as Fig.\ref{fig:1} but for $^{\rm 272}$Ds.}
\label {fig:8}
\end{figure}

\begin{figure}
\includegraphics[width=15cm]{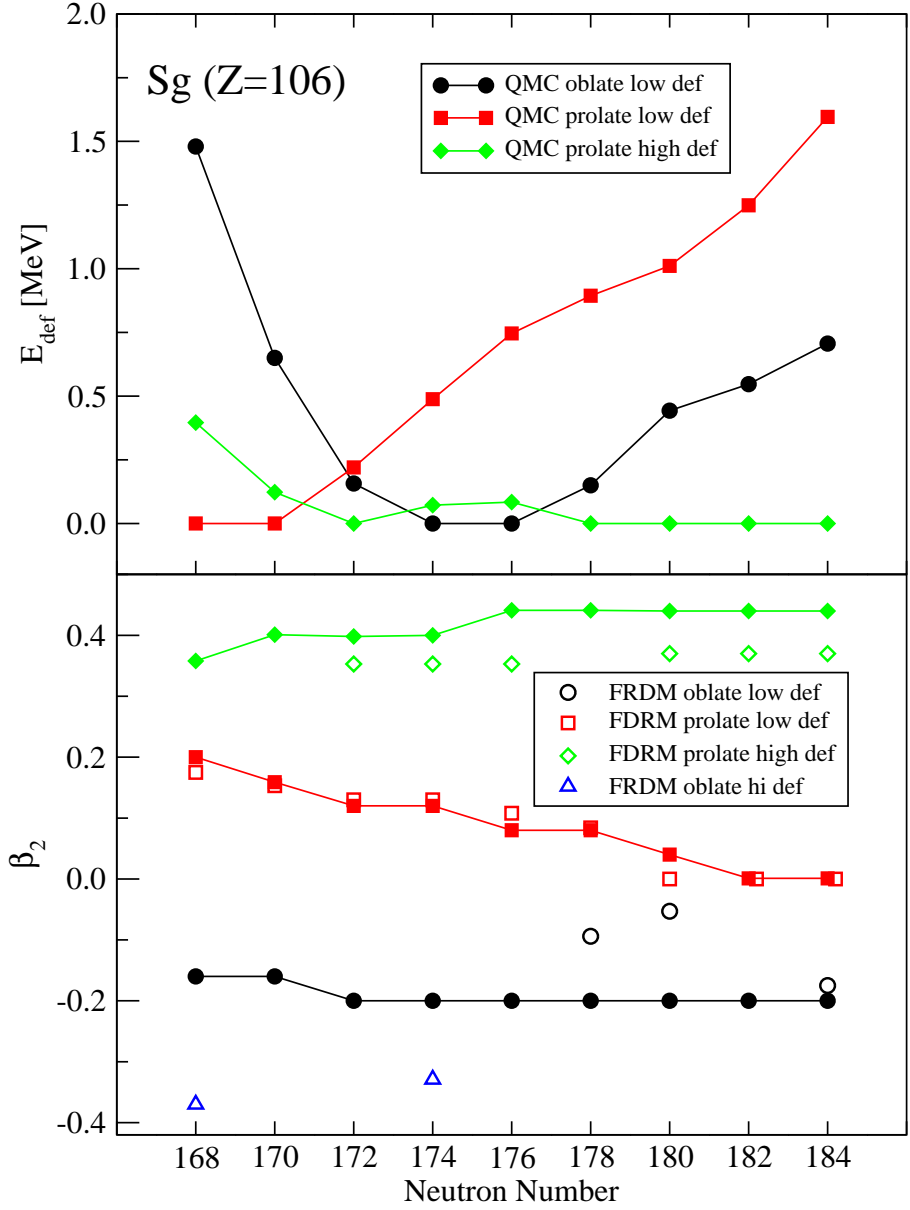}
\centering{}\caption{Shape coexistence in heavy Sg isotopes. $E_{def}$ of coexisting shapes are shown in the top panel as a function of the neutron number. Corresponding deformation parameters $\beta_2$ are in the bottom panel (full symbols). Results from FDRM \cite{moller2012} are added for comparison (empty symbols). The label `low def' refers to $\mid\beta_2\mid  \le 0.20$, `high def' stands for $\mid\beta_2\mid  \sim 0.40 $. See text for more discussion.}
\label {fig:9}
\end{figure}

The QMC model predicts an intriguing evolution of shapes, including shape coexistence, prolate-oblate and oblate-spherical shape changes as a function of neutron and proton numbers. We illustrate this behaviour in detail in Figs.~\ref{fig:1}--\ref{fig:8}. The deformation energy $E_{def}$, defined as the $E^{cal}_{b}(Z, N) - E^{eq}_{b}(Z, N)$ and normalized to $E^{eq}_{b}(Z,N)$ at the minimum,  is plotted as a function of $\beta_{\rm 2}$ for  Cm, Cf, Fm, No, Rf, Sg, Hs and Ds isotopes with neutron number 138 $\leq$ N $\leq$ 184, respectively.  Examination of the figures reveals several interesting results: \\
\noindent
(i) a well developed prolate deformation with  $\beta_{\rm 2} \sim$ 0.28 is predicted for all isotopes with neutron number 138$\leq$ N $\leq$ 152,\\
\noindent
 (ii) this deformation decreases for neutron numbers 154 $\leq$ N $\leq$ 166, \\
\noindent
 (iii) isotopes of Cm, Cf, Fm and No undergo a prolate-oblate transition at N=168, an oblate-spherical transition at N=178 and remain spherical until N=184 as expected, \\
\noindent
(iv) isotopes of Rf undergo a prolate-oblate transition at N=170 and remain oblate for all N up to N=184, \\
\noindent
(v) isotopes of Sg develop an interesting shape coexistence and shape transitions for N $\ge$ 168, illustrated in Fig.~\ref{fig:9} and discussed more in detail in the following paragraph, \\
\noindent
(vi) isotopes of Hs and Ds remain prolate for N $\ge$ 168 and, for N=172 up to N=184. become highly deformed with $\beta_{\rm 2} \sim$ 0.4.

Comparing data available in tabulations \cite{moller2016, muntian2001,muntian2003,muntian2003a}, the values of  $\beta_{\rm 2} $ in the region of neutron numbers 138 $\leq$ N $\leq$ 152 calculated in QMC, FRDM and MM are all very similar. The differences set in for N $>$ 154.  FRDM predicts for all nuclei studied in this work a prolate-oblate shape transition preceding development of a spherical shape at N=184 at N=175 (Cm), N=176 (Cf and Fm), N=177 (No and Rf) and N=178 (Sg, Hs and Ds). The MM model indicates a decrease in quadrupole deformation with N increasing towards N=184, reaching spherical shapes at N$\sim$180 but does not predict any prolate-oblate transitions.

Extensive FRDM and FRLDM calculation of ground-state potential energy surfaces leading to identification of nuclear shape isomers has been published in \cite{moller2009,moller2012}. The \cite{moller2012} calculation has been performed in 3D space of $\epsilon_2, \epsilon_4$ and $\gamma$ deformation parameters. The tabulation provides data on coexisting potential energy minima and saddles between them for 7206 nuclei from A = 31 to A = 290. Numerical data are provided in tables for all nuclei with N $\leq$ 260. Results for nuclei closer to the neutron drip-line are presented as figures. These data can be used for a  qualitative comparison with the QMC results obtained in a 2D space with $\beta_2$ and $\beta_4$ deformation parameters, assuming axial and time-reversal symmetry.

Taking Sg isotopes in the range of neutron numbers N $>$ 168 as an example, we show in the top panel for Fig.~\ref{fig:9} the coexistence of different shapes up to $E_{def} \sim$ 1.5 MeV. The corresponding values of $\beta_2$ are given in the bottom panel (full symbols). We observe a relatively slow change in deformations but a distinct, systematic change in the deformation energy with increasing neutron number. The QMC results are qualitatively compared with FRDM predictions \cite{moller2012} obtained from figures for $\gamma$=0 (taking roughly $\beta_2 \sim$ 1.06 $\epsilon_2$). It is interesting to notice that the most of deformed shapes predicted by QMC appear also in the FRDM model. However no attempt has been made to compare deformation energies. The major difference is that QMC predicts the disappearance of the N=184 spherical shell gap, seen in the FRDM and MM models for Rf, Sg, Hs and Ds. 

\subsection{Signatures of shell gaps based on bulk properties}
\label{search_bulk}
It is intuitive that energy gaps between the single-particle levels of heavy deformed nuclei will become smaller than in lighter nuclei. Energy levels with higher angular momentum become close the Fermi surface and the level density increases. The attempts to detect energy gaps in a single-particle spectrum may be hindered by the limited accuracy of individual binding energies calculated in theoretical models. In this section we discuss the performance of the QMC model in detecting fingerprints of  the N=152 and 162 deformed shell gaps, predicted in macro-micro models, as detailed in Secs. ~\ref{s2n}, \ref{delta2n}, \ref{qalpha} and \ref{pairing}.

\subsubsection{Two-neutron separation energies S$_{\rm 2n}$}
\label{s2n}
The location and size of shell gaps can be extracted from two-neutron separation energies S$_{\rm 2n}$(Z,N)=E$_{\rm b}$(Z,N) - E$_{\rm _b}$(Z,N-2), where a shell gap is visible through a sharp drop as a function of neutron number.This drop reflects the increased change in Fermi energy when crossing the energy gap in the single-particle spectrum. Current experimental data compiled in AME2016 provide clear evidence for the N=162 gap and only a weak indication for the N=152 gap for nuclei with atomic number between Z=98 and Z=110  (see Fig.4 in \cite{block2015}). 

\begin{figure}
\includegraphics[width=18cm]{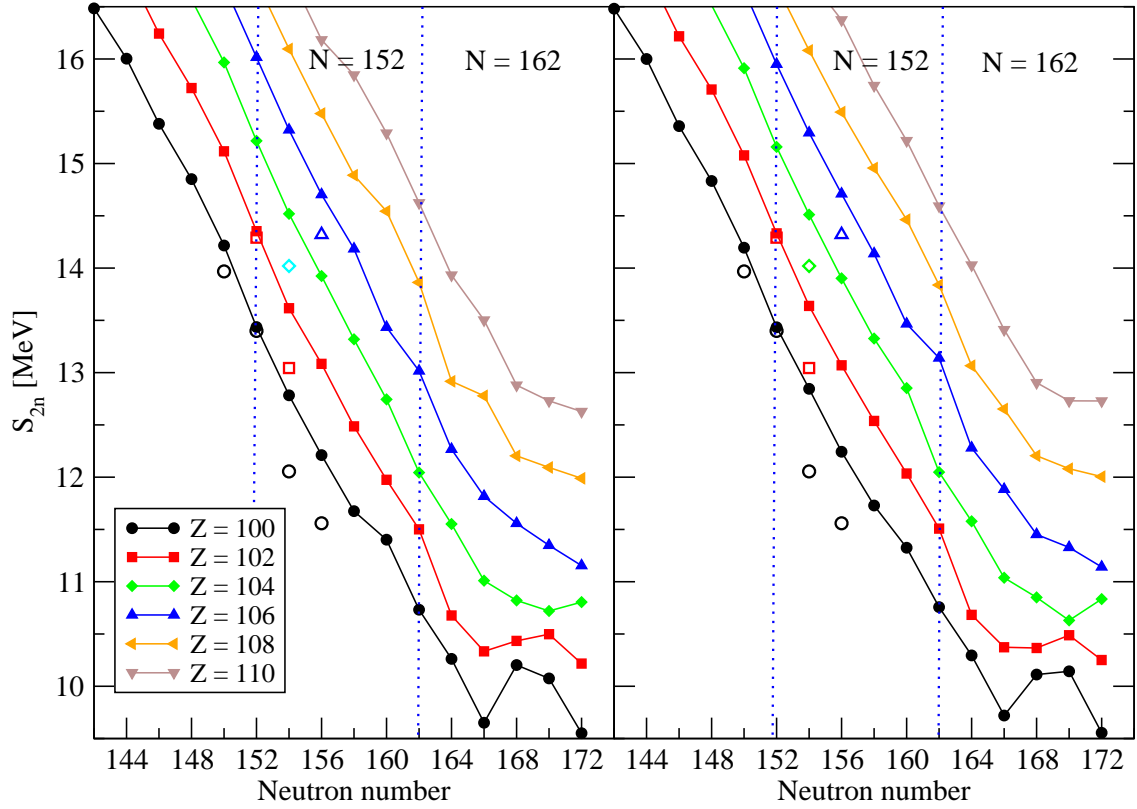}
\centering{}\caption{Two-neutron separation energies as calculated in the QMC model using the mesh (right panel) and the fit method (left panel). The dotted blue lines indicate N=152 and N=162, where the shell gaps are expected. Experimental data for the same range of proton and neutron numbers are taken from  AME2016 \cite{wang2017} and also  can be found in Fig.4 of \cite{block2015}. }
\label {fig:10}
\end{figure}

We illustrate results of the QMC calculation Fig.~\ref{fig:10}. Data obtained in the mesh calculation (see Sec.~\ref{calmeth}) are displayed in the left panel and those with the fit correction in the right panel. The vertical dashed lines indicate the neutron numbers N=152 and N=162. Experimental data taken from AME2016 are indicated with crosses. The QMC data predict systematically a faster decrease with increasing neutron number. The total  RMS deviation of the QMC result from experiment is 0.470 MeV.  No evidence for a change in the slope S$_{\rm 2n}$ can be found at N=152 in either panel. A weak enhancement of the S$_{\rm 2n}$ slope appears at and in the vicinity of N=162. Interesting variation of  S$_{\rm 2n}$ with neutron numbers 168 $\leq$ N $\leq$ 172, appearing in the experimental data as well but within large errors \cite{block2015}, is most likely  related with prolate-oblate shape changes and coexistence discussed in Sec.~\ref{gd}. It is important to realize that ground state binding energies and their differences reflect not only shell structure but also other factors, such as deformation and its changes.

\subsubsection{Empirical shell-gap parameter $\delta_{\rm 2n}$}
\label{delta2n}
The sensitivity to deformed shell gap is enhanced in the empirical neutron shell gap parameter $\delta_{2n}$,  the difference between two-neutron separation energies \cite{bender2003}
\begin{equation}
 \delta_{\rm 2n} = S_{\rm 2n}(Z,N)-S_{\rm 2n}(Z,N+2)=2E_{\rm b}(Z,N)-E_{\rm b}(N-2)-E_{\rm b}(Z,N+2)
\end{equation}
with E$_{\rm b)}$ being the ground state binding energy of a single isotope. The quantity $\delta_{\rm 2n}$ shows maximum at the nucleon number for which a (sub)shell gap occurs. This differential quantity amplifies the visibility of weak shell effects and takes advantage of \textit{differences} rather than single values of binding energies. However, it works only if \textit{no dramatic rearrangement} in the mean field takes place between the three adjacent nuclei, (Z, N-2), (Z,N) and (Z,N+2) \cite{bender2003}. It is also important to note that other factors, such as a change in deformation and/or pairing, influences the value of $\delta_{\rm 2n}$.
 
Block \cite{block2015} studied the N=152 shell gap in $^{\rm 150-158}$No  isotopes to illustrate the importance of high precision direct mass measurements in the quest for regions of stability in the SHE region. In their Fig.~5 the quantity  $\delta_{2n}$ was shown in comparison with experiment and predictions of the FRDM and MM models and mean-field Hartree-Fock-Bogoliubov models, showing clearly the failure of the latter. 

\begin{figure}
\includegraphics[height=17cm, width=15cm]{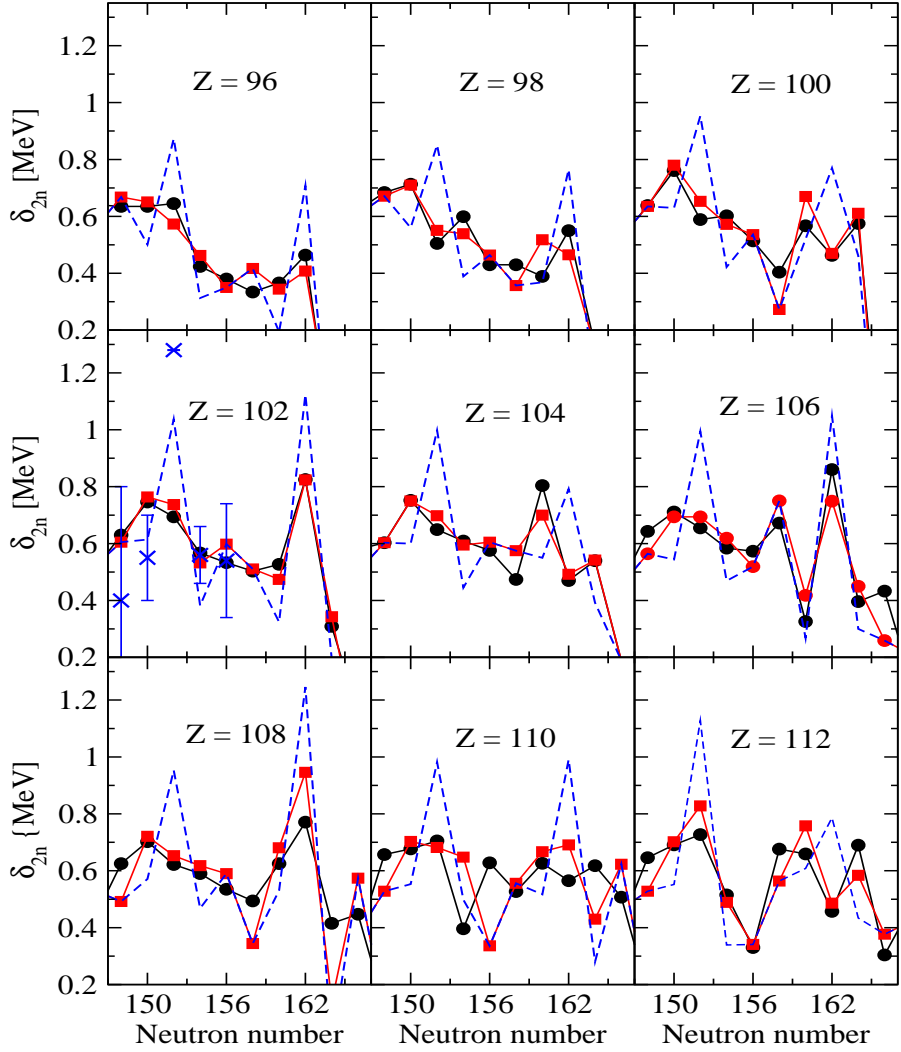}
\centering{}
\caption{Empirical  shell gap parameter $\delta_{2n}$ for 96 $\leq$ Z $\leq$ 112 and 148 $\leq$ N $\leq$ 166 as calculated in QMC. The full black circles correspond to the mesh calculation and the full red squares illustrate results with the fit calculation. Full blue triangles, appearing in the panel for Z = 102 represent experimental data taken from \cite{wang2017,block2015}. The dashed blue curve shows $\delta_{2n}$ values with the binding energies of isotopes with N=152 and 162 artificially increased by 150 keV in all elements. For discussion see the text.}
\label{fig:11}
\end{figure}

The parameter $\delta_{2n}$ in the QMC model was calculated again using both mesh and fit method and the results are displayed in Fig.~\ref{fig:11}. Similarly to the two-neutron separation energies, no clear evidence was found for the N=152 deformed shell gap.
Experimental data taken from \cite{block2015} for Z=102 were added to the corresponding panel in Fig.~\ref{fig:11}. It can be seen that the calculation agrees with the experiment within errors for N=148, 150, 154 and 156 but fails by more than the factor of two  to predict the enhancement at N=152.  There is some enhancement at N=162 for Z=102, 106 and 108 but its magnitude would need to be tested when experimental data become available. In order to understand better the requirement on precision of ground state binding energies used to calculate $\delta_{2n}$ we artificially increased the binding energy of all isotopes with N=152 and 162 by 150 keV (which is less than 0.01\%), leaving all the other binding energies the same. The result is illustrated in all panels of Fig.~\ref{fig:11} by the thin dashed blue curve, showing clear maxima at N=152 and 162. Most mean-field models based, in particular, on traditional Skyrme interactions, do not have the accuracy to reveal such weak shell closures. However, for example, the recent work by the Goriely group \cite{goriely2016} reported a mass model HFB-31 which involves  23 parameters and claims \textit{model} accuracy 0.561 MeV.  Unfortunately the on-line available mass table \cite{brussels} does not include the latest HFB-31 results, but the last accessible HFB-29 model yields for $^{252,254,256}$No the difference between the calculated and experimental mass excess -0.19, -0.31 and -0.08 MeV respectively, which is encouraging.

\subsubsection{Q$_\alpha$ values}
\label{qalpha}
The $\alpha$- decay life-times, T$_{\rm 1/2}$, are exponential functions of the energy release, Q$_\alpha$(Z,N) = $E_b(Z-2,N-2)$ - $E_b(Z,N)$ +$E_b(2,2)$ in the decay, which, in turn, depends on the mass (binding energy) difference between the parent and daughter states and the binding energy of the $\alpha$ particle, $E_b(2,2)$, taken as 28.3 MeV. Thus, while the absolute values of the nuclear masses are not crucial in this context, the differences are essential. For instance, a change in Q$_\alpha$ by 1 MeV in a nucleus with Z=118 would make a difference in T$_{\rm 1/2}$ of three orders of magnitude \cite{stone2017}. It is therefore desirable to aim for as accurate as possible predictions of  Q$_\alpha$ to provide a useful guide for experiment. The relation between Q$_\alpha$ values and shell gaps has been established on experimental grounds, as mentioned in Sec.~\ref{intro}.

\begin{figure}
\includegraphics[width=18cm]{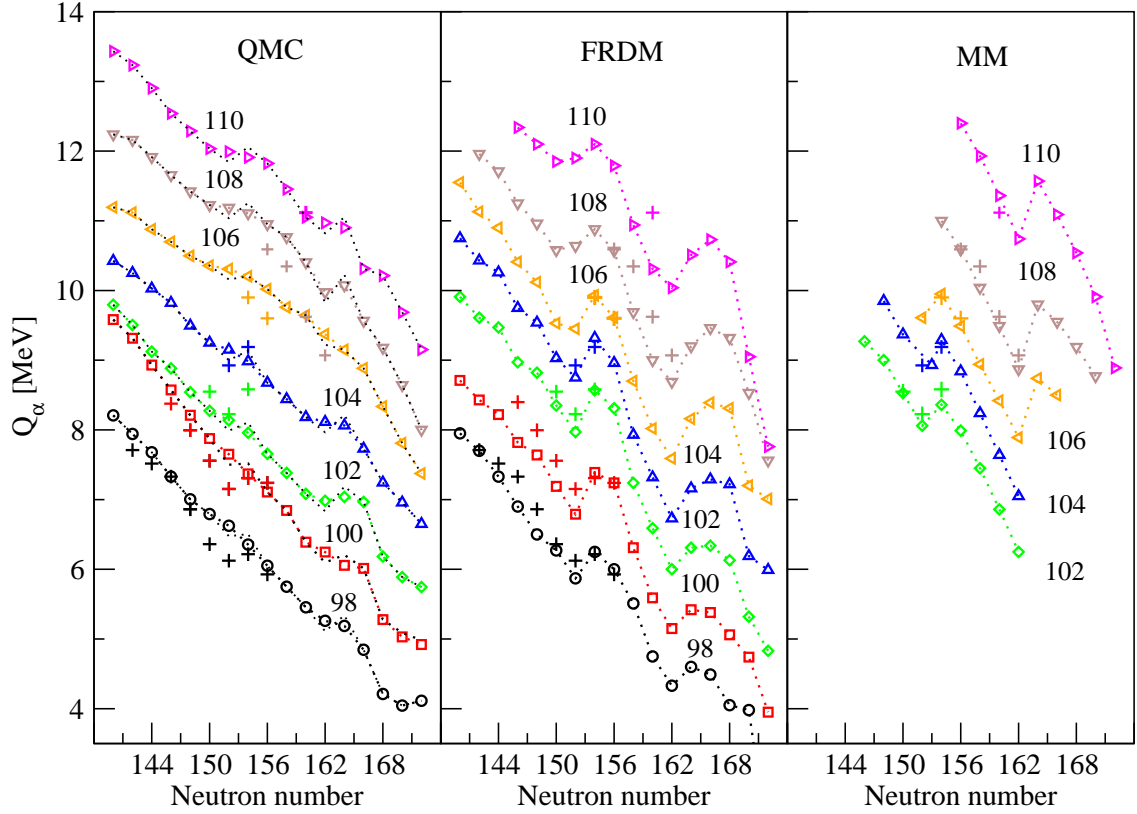}
\centering{}\caption{ Q$_\alpha$(Z,N)  for isotopes of SHE with 98 $\leq$Z$\leq$ 110 and 148 $\leq$N$\leq$ 170 as calculated in the QMC, FRDM and MM models. Experimental data, depicted by crosses in the same colour as the calculated points, were taken from AME2016 \cite{wang2017}. All entries are in MeV.}
\label{fig:12}
\end{figure}

The QMC results for Q$_\alpha$(Z,N)  for isotopes of SHE 98 $\leq$Z$\leq$110 with 148 $\leq$N$\leq$ 170 as calculated in the QMC, FRDM and MM models, together with the available experimental data are shown in Fig.~\ref{fig:12}. The macro-micro models predict a significant decrease in  Q$_\alpha$ value at N=152 and 162. This decrease agrees with available (sparse) experimental data reasonably well although the effect is somewhat over-estimated in FRDM. The MM model shows the best agreement with data, as in the case of absolute ground state binding energies, because of its special adjustment to the SHE region.

The QMC model shows limited sensitivity of the  Q$_\alpha$(Z,N) to the shell gap effects predicted by the FRDM and MM models, in particular at N=152. The effect of the N=162 shell gap is predicted in the No and Hs isotopes, with a weakly increased tendency toward stability in Fm, Rf and Ds. No N=162 effect has been shown in Sg nuclei.

\section{Pairing}
 \label{pairing}
The contribution of the residual pairing interaction to the total ground state energy increases with increasing density of single-particle states around the Fermi surface.  The presence of shell gaps invokes two opposing effects, the decrease of pairing and the increase in nuclear stability. In this way, a decrease in pairing energy would signal the existence of a shell gap.

\begin{figure}
\includegraphics[width=16cm,height=18cm]{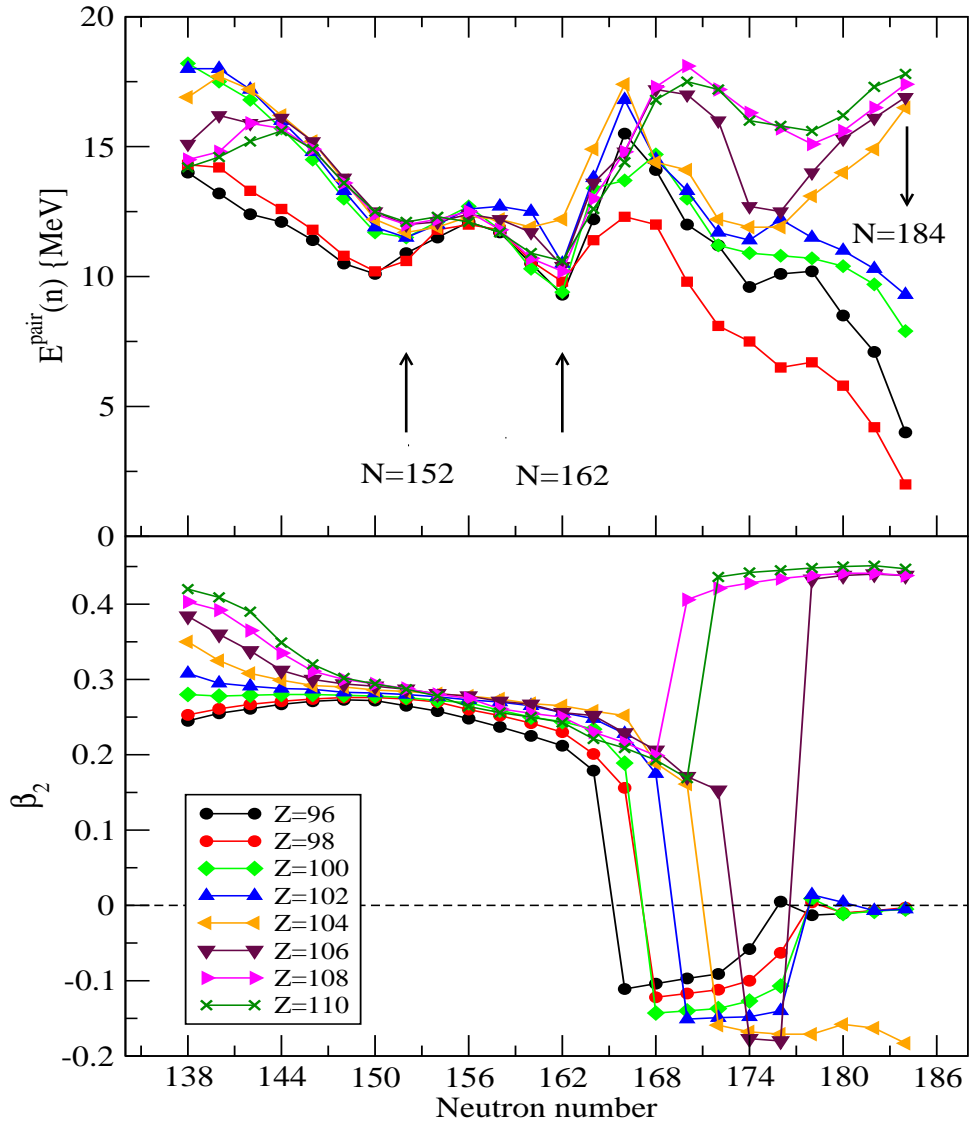}
\centering{}\caption{Neutron pairing energy as a function of neutron number (top panel). The quadrupole deformation parameter $\beta_{\rm 2}$ is shown in the bottom panel. The (black) dashed line indicate a  spherical shape. More explanation can be found in the text.}
\label{fig:13}
\end{figure}

We show in Fig.~\ref{fig:13} the neutron pairing energy as a function of neutron number N, 138$\leq$ N $\leq$184 for 98 $\leq$ Z $\leq$110 (top panel) as a function of quadrupole deformation (bottom panel). Because the pairing energy depends on the single-particle spectrum, mainly close to the Fermi surface, it follows that it is dependent on deformation. Examination of the figure shows that QMC model predicts the shell gap around N=152, most strongly in Cm and Cf, and its weakening for  heavier elements. The dip at N=162 is present for all Z and grows with increasing Z.  It is, however, close to the region where shape changes start to take place (see Sec.~\ref{gd}) and is most likely influenced by their presence. Another remarkable feature is the prediction of disappearance of the spherical N=184 shell gap in Rf, Sg, Hs and Ds isotopes. The competition between the spherical ground state and the newly developed highly deformed state with $\beta_{\rm 2} \sim 0.4$ in its close vicinity in neutron heavy isotopes with N$\ge$170 is most likely responsible for this effect and the pairing energy provides further indication for it taking place.  It is clearly seen that the neutron pairing energy is decreasing with neutron number  increasing N = 184 in Cm, Cf, Fm and No but is increasing in Rf, Sg, Hs and Ds nuclei. This trend is consistent with increasing level density around the Fermi surface in the latter group, inconsistent with the presence of an energy gap and reduced pairing.

\section{Single-particle spectra and shell gaps}
\label{sps}
In the previous sections we demonstrated that bulk ground state properties of isotopes of SHE elements, as calculated in the QMC model, have limited sensitivity to shell gaps. In this section we turn to a direct signature of energy gaps obtained from single-particle spectra themselves. 

As already mentioned in Sec.~\ref{calmeth}, the ``single-particle'' energies obtained in mean-field models  are eigenvalues of the model Hamiltonian and as such are model dependent. The theoretical single-particle states calculated in the ground state of even-even nuclei can be approximately identified with experimental data only if the modifications of the mean field, induced by the extra nucleon (or hole), are small \cite{bender2003}. In the vicinity of doubly magic nuclei these modifications can be reduced to polarization effects. In open-shell deformed nuclei the modifications are hard to account for as they include, amongst other things, changes in pairing fields and, in case of odd-A and odd-odd nuclei, breaking of time-reversal symmetry.

There is also an uncertainty in spin-parity assignments to experimental levels of odd-A and odd-odd nuclei.  The assignment is often made by a comparison with the predictions of a shell or the Nilsson model \cite{nilsson1969} or even a simple rotation model with a residual interaction \cite{chasman1977}. More recently, large scale shell model calculations are also used but they are also model dependent. Alternatively one may simply be guided by empirical systematics, which may significantly propage one mistaken assignment through a chain of nuclei.  In SHE region, the spin-parity assignments are often deduced from $\alpha$ and $\beta$ decay chains. Unless at least one member of the chains is known from independent measurement of electromagnetic moments and/or transition probabilities, the assignments may be ambiguous.

Keeping in mind the complexity of both calculated and experimental single-particle energies and their relationship, we examine the performance of the QMC model in predicting single-particle states, first, in Sec.~\ref{light}  for lighter nuclei below and including $^{208}$Pb where experimental data exist, followed by calculations in the SHE region in Sec.~\ref{heavy}, where experimental data are scarce.

\subsection{Nuclei with Z$\leq$82 and N$\leq$126}
\label{light}

Single-particle spectra studied in this subsection, calculated for (semi)magic nuclei $^{40}$Ca, $^{48}$Ca, $^{56}$Ni,  $^{56}$Ni , $^{78}$Ni , $^{90}$Zr, $^{100}$Sn , $^{132}$Sn, $^{146}$Gd  and $^{208}$Pb with a (semi)magic number of protons and neutrons, include bound states both below and above the Fermi level of each nucleus.

 Experimental data leading to spin-parity assignment of single-particle states, supported by spectroscopic factors obtained in single nucleon transfer reactions seem to be relatively safe \cite{grawe2007} and may be used for most states in $^{16}$O, $^{40}$Ca, $^{48}$Ca and $^{208}$Pb.  For $^{132}$Sn the lowest particle (hole) levels are associated with single-nucleon states in adjacent odd-A nuclei $^{131}$In and $^{131}$Sn , while the $^{100}$Sn and $^{78}$Ni values were extrapolated from regions of stable nuclei by shell model calculations. Data for $^{56}$Ni were inferred from a radioactive target experiment  (see \cite{grawe2007} and references therein for details). Typical errors on the single-particle energies are around 100 keV, although in some cases, such as $^{100}$Sn, they may reach between 300 and 500 keV \cite{grawe2001}. Errors reported on single-particle states in $^{90}$Zr, also obtained in single-particle transfer reactions, are even larger, in some cases over 1 MeV \cite{bespalova2005}. 

\begin{figure}
\includegraphics[width=18cm]{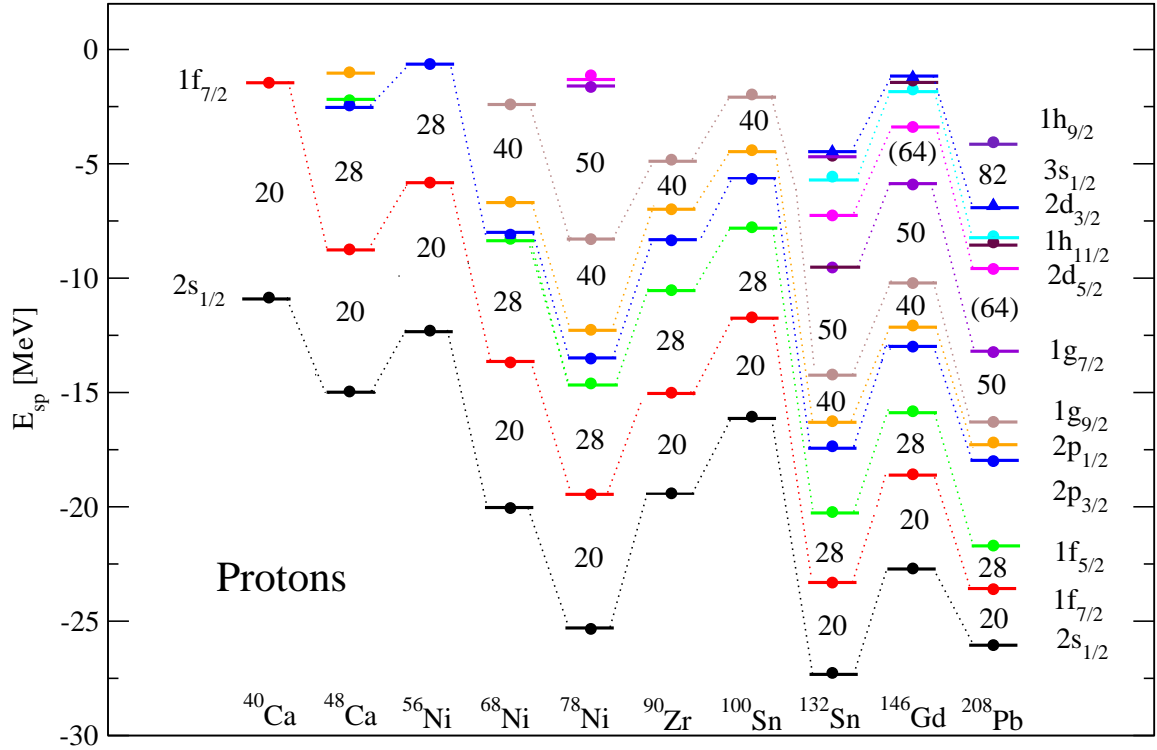}
\centering{}\caption{Proton single particle energies for $^{\rm 40}$Ca, $^{\rm 48}$Ca, $^{\rm 56}$Ni, $^{\rm 68}$Ni, $^{\rm 78}$Ni, $^{\rm 90}$Zr, $^{\rm 100}$Sn, $^{\rm 132}$Sn, $^{\rm 146}$Gd and $^{\rm 208}$Pb. For more explanation see the text.}
\label{fig:14}
\end{figure}

\begin{figure}
\includegraphics[width=18cm]{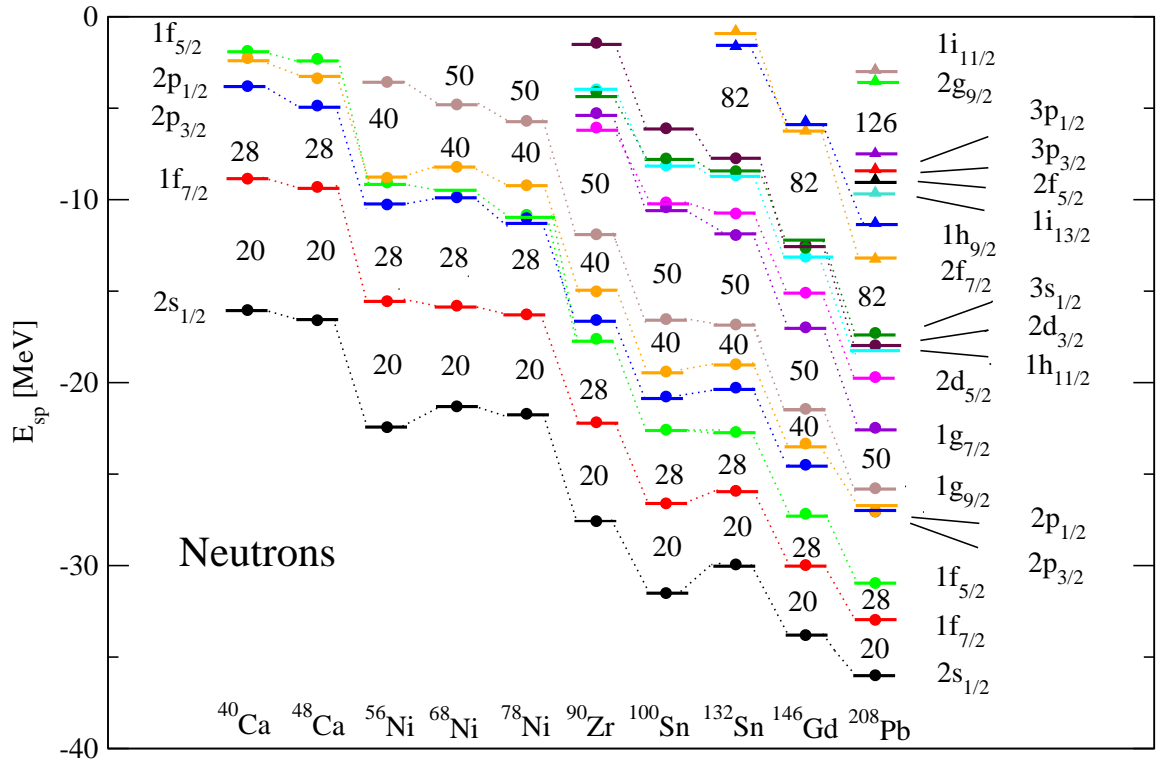}
\centering{}\caption{The same as Fig.~\ref{fig:14} but for neutron single-particle states.}
\label{fig:15}
\end{figure}

The proton and neutron single-particle spectra are shown in Figs.~\ref{fig:14} and \ref{fig:15} and a comparison with experimental data in Tables~\ref{tab:2} and \ref{tab:3}. It is gratifying to see that all the major proton and neutron (sub)shell closures are reproduced in the calculation and the sequence and values of the single-particle energies are, in most cases, in a good agreement with experiment.

 For instance, the Z=28 shell gap is formed between $1f_{7/2}$ and  $2p_{3/2}$ or  $2f_{5/2}$. The ground state spin and parity of odd-proton ground states of Cu(Z=29) are known to be 3/2$^{-}$ for N=28 to N=44. Spin-flip to 5/2$^-$ has been identified \cite{koester2011,flanagan2009} at N=46.  Furthermore, it is interesting to note the recent mass measurement of $^{75-79}$Cu \cite{welker2017}, which supported the doubly magic character of $^{78}$Ni. These data yield the proton Z=28 shell gap to be 6.7 MeV in  $^{68}$Ni and to reduce to 4.9 MeV at $^{78}$Ni. The QMC model predicts the gaps between $1f_{7/2} \rightarrow 1f_{5/2}$ 5.4 and 4.8 MeV, respectively, with a somewhat smaller reduction. Spectroscopic study of $^{79}$Cu \cite{olivier2017} brought evidence that the Z=28 subshell gap is still active for N=50. The results confirmed that the $^{79}$Cu nucleus can be described in terms of a valence proton outside a $^{\rm 78}$Ni core, thus producing indirect evidence of the magic character of the latter. The QMC model supports these conclusions.

In the neutron spectrum, the experimental ground state of $^{\rm 131}$Sn has spin 3/2  while all mean-field models, Skyrme Hartree-Fock, Gogny, and relativistic mean-field, predict spin 11/2 \cite{bender2003} . The QMC model makes a correct prediction of spin 3/2 for the first hole state in $^{\rm 132}$Sn, as demonstrated in Fig.~\ref{fig:15}. Other examples may be found in measurements of magnetic moments, (see e.g. Ref.~\cite{rikovska2000}). Precise measurement of the ground state magnetic moment of $^{67}$Ni revealed close agreement with the value expected for the 2p$_{\rm 1/2}$ hole in the doubly (semi) magic nucleus $^{68}$Ni (Z=28, N=40). As seen in Fig.~\ref{fig:15}, the 2p$_{\rm 1/2}$ state is the first orbital located under the Fermi level in $^{68}$Ni predicted by the QMC model. 

More detailed discusion of shell gaps in the region below $^{208}$Pb is beyond the scope of this paper but can be found in Ref.~\cite{sorlin2008}.

\subsection{Nuclei with 96$\leq$Z$\leq$110 and 148$\leq$N$\leq$162}
\label{heavy}

Single-particle spectra of SHE isotopes are particularly sensitive to details of the shell structure and the model prediction of spin-orbit splitting. The level density around the Fermi surface increases as compared with lighter nuclei, the shell gaps are smaller and their experimental and theoretical more involved.

In this section we investigate neutron single-particle energies in $^{244}$Cm,  $^{248}$Cf, $^{252}$Fm, $^{256}$No, $^{260}$Rf, $^{264}$Sg, $^{268}$Hs and $^{272}$Ds, as representative examples of evolution of the shell structure with deformation parameter 0 $\leq \beta_{\rm 2} \leq$ 0.55 in the SHE region which is likely to be accessible to experiment.  Oblate shapes are not considered because, as demonstrated in Sec.~\ref{gd}, they do not predicted to occur in these isotopes.

\begin{figure}
\includegraphics[width=18cm]{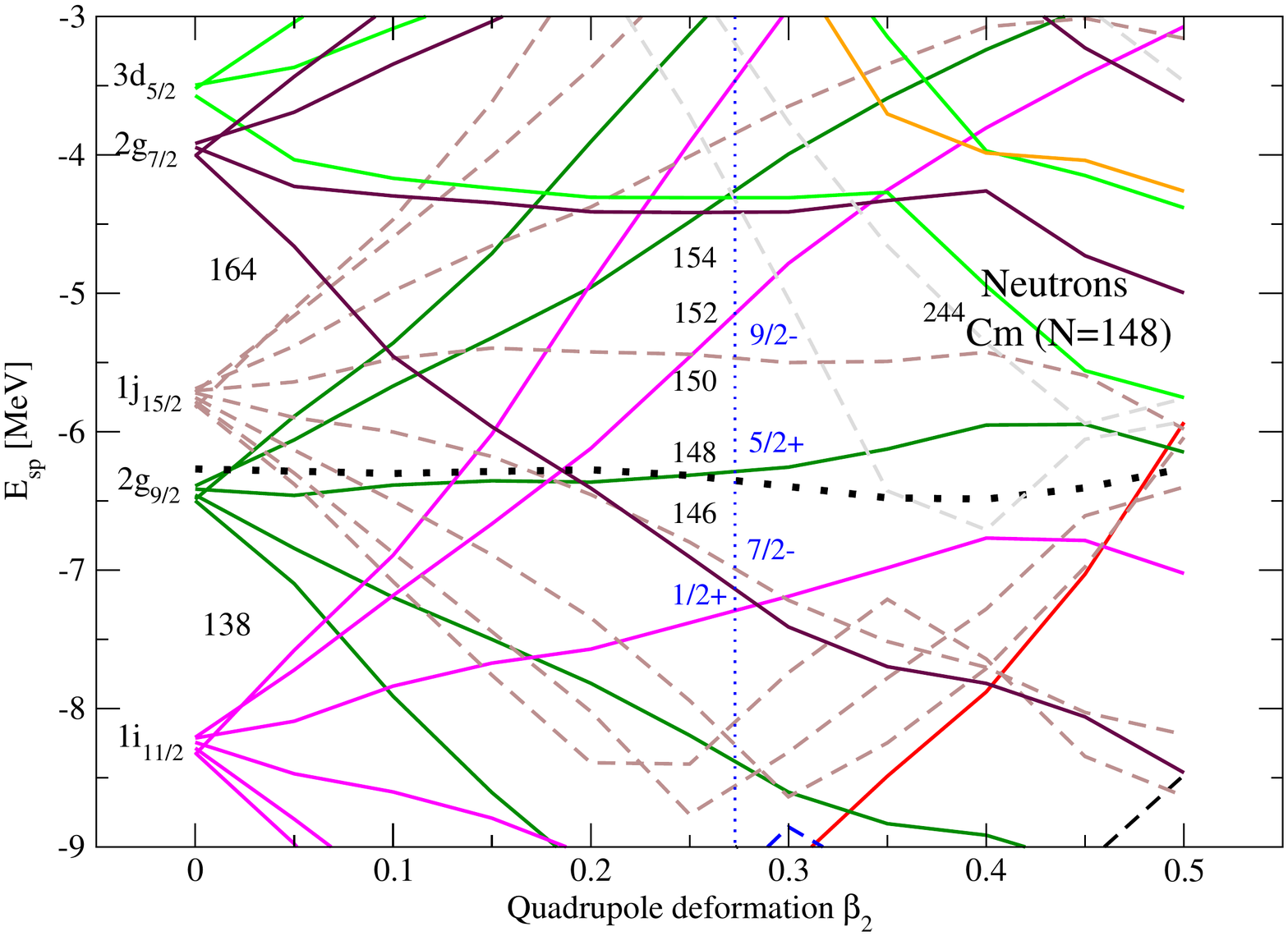}
\centering{}\caption{Neutron single-particle orbitals in $^{\rm 244}$Cm as a function of quadrupole deformation as calculated in the QMC model. Negative parity orbitals are depicted by dashed lines. The Fermi surface is indicated by a thick (black) dotted line.The thin (blue) dotted vertical line indicates the ground state deformation parameter $\beta_{\rm 2}$. $\Omega^\pi$ values have been added to orbitals close to the Fermi surface. For more discussion see the text.}
\label{fig:16}
\end{figure}

\begin{figure}
\includegraphics[width=18cm]{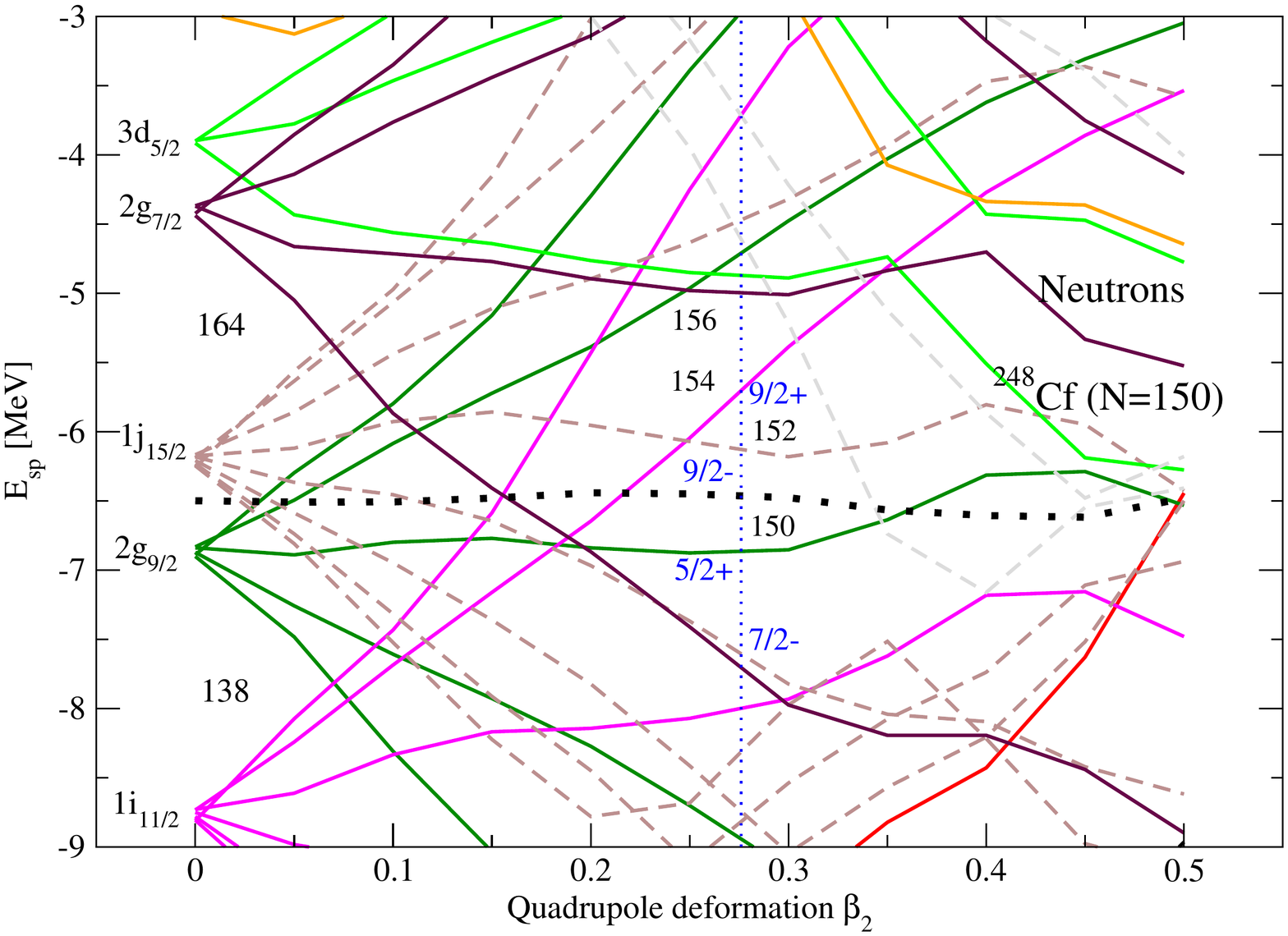}
\centering{}\caption{The same as Fig.~\ref{fig:16} but for $^{\rm 248}$Cf.}
\label{fig:17}
\end{figure}

\begin{figure}
\includegraphics[width=18cm]{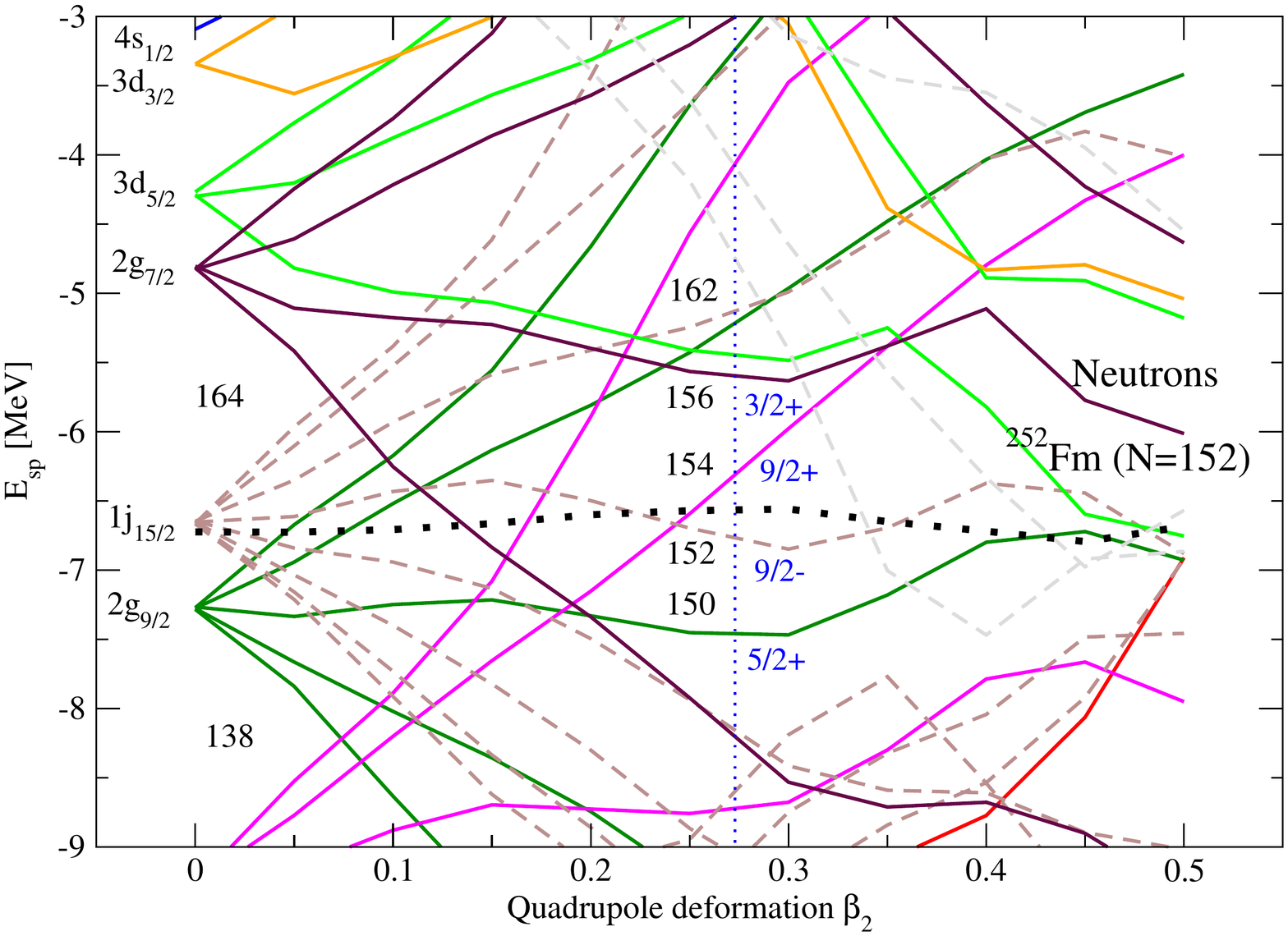}
\centering{}\caption{The same as Fig.~\ref{fig:16} but for $^{\rm 252}$Fm.}
\label{fig:18}
\end{figure}

\begin{figure}
\includegraphics[width=18cm]{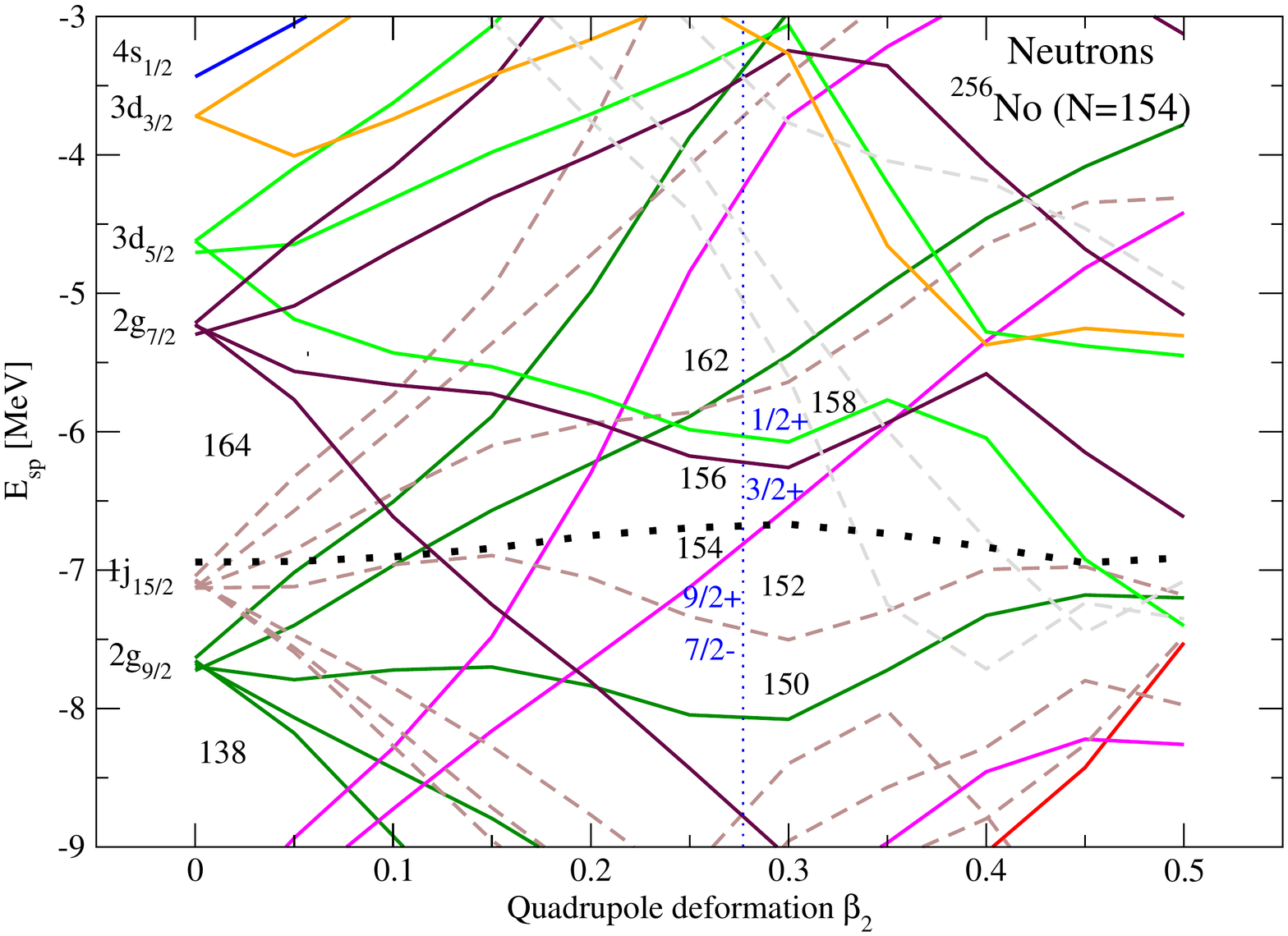}
\centering{}\caption{The same as Fig.~\ref{fig:16} but for $^{\rm 256}$No.}
\label{fig:19}
\end{figure}

\begin{figure}
\includegraphics[width=18cm]{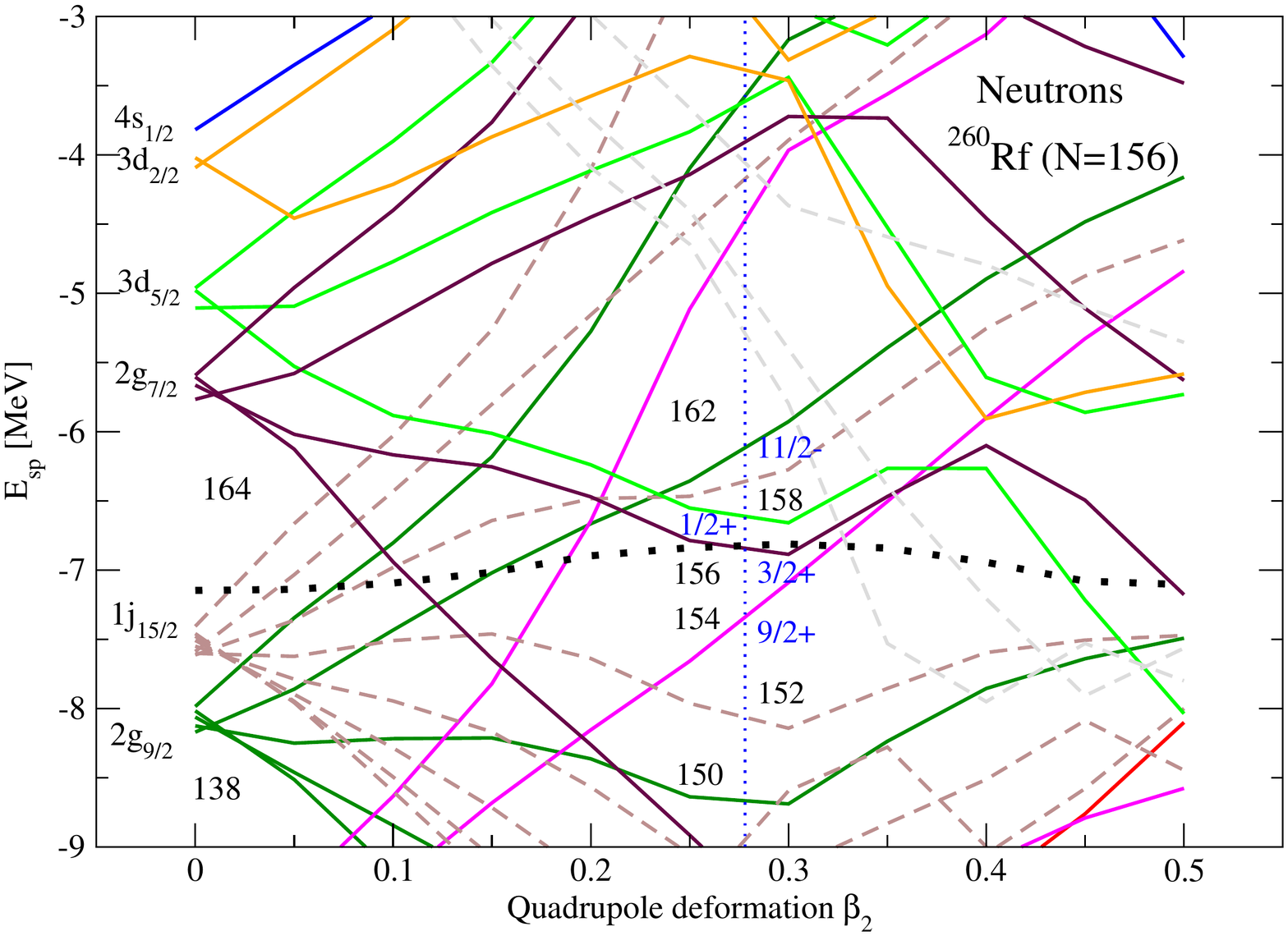}
\centering{}\caption{The same as Fig.~\ref{fig:16} but for $^{\rm 260}$Rf.}
\label{fig:20}
\end{figure}

\begin{figure}
\includegraphics[width=18cm]{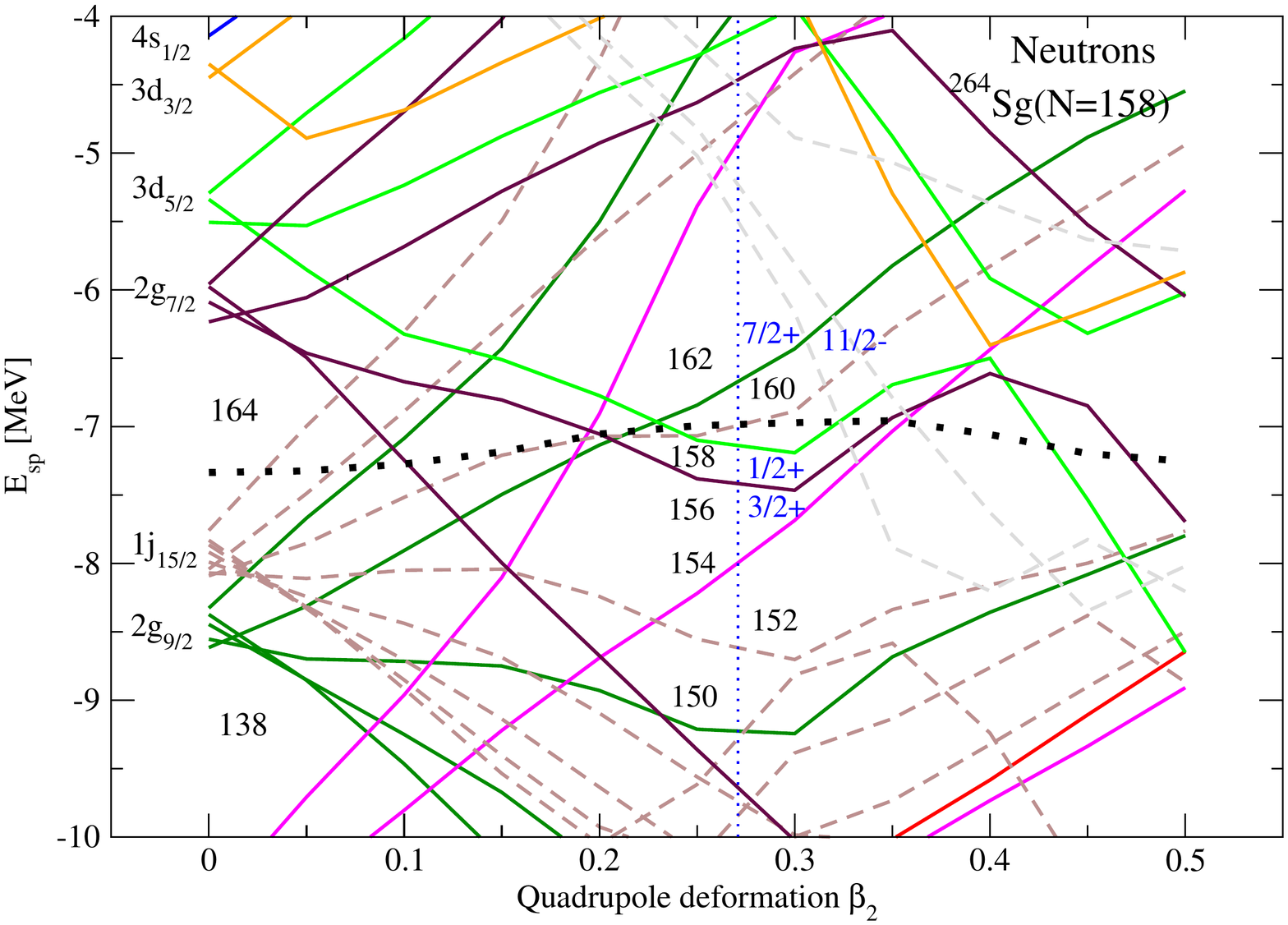}
\centering{}\caption{The same as Fig.~\ref{fig:16} but for $^{\rm 264}$Sg. Note that the y-axis was displaced by 1 MeV without a change of scale.}
\label{fig:21}
\end{figure}

\begin{figure}
\includegraphics[width=18cm]{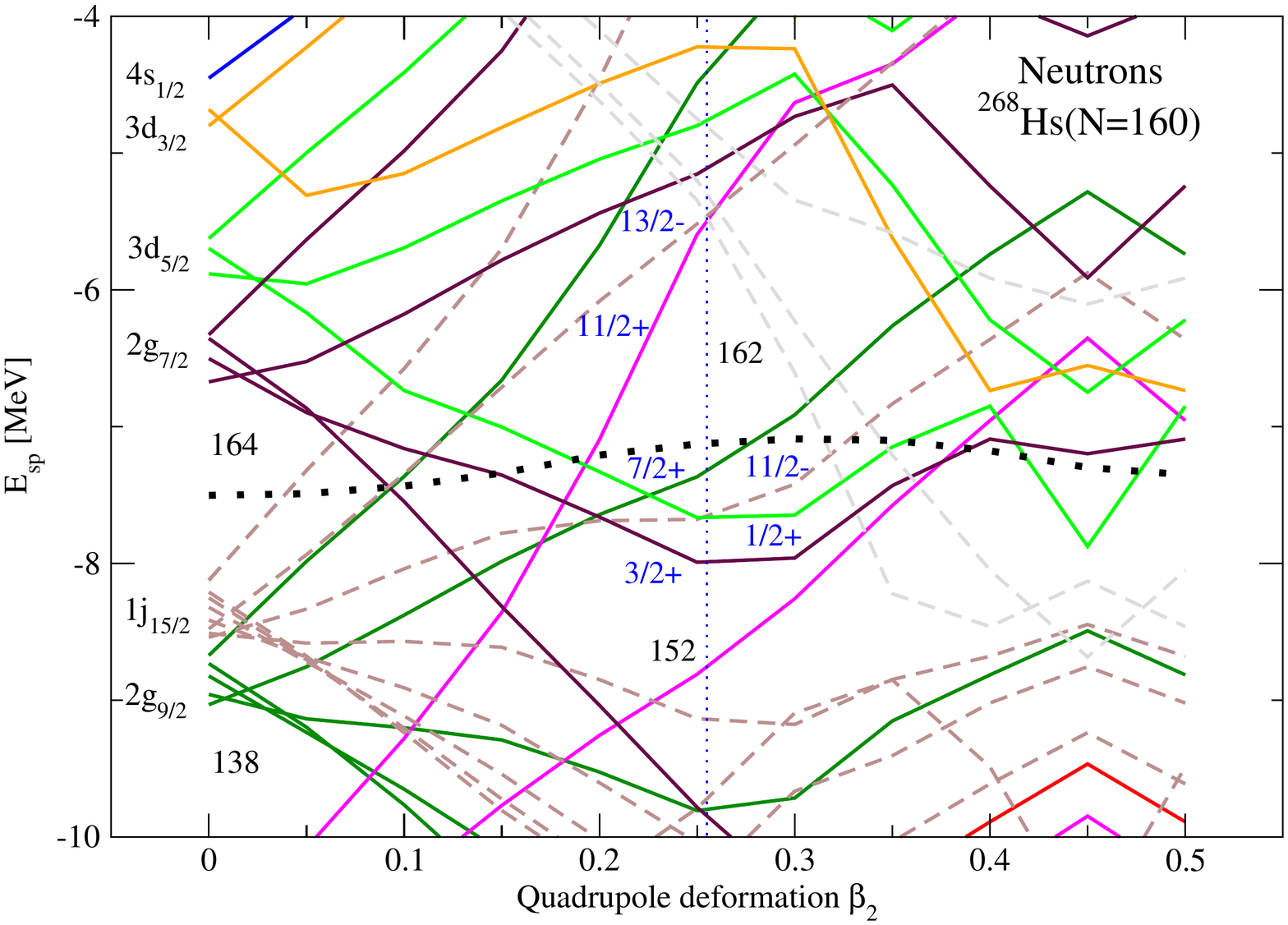}
\centering{}\caption{The same as Fig.~\ref{fig:16} but for $^{\rm 268}$Hs. Note that the y-axis was displaced by 1 MeV without a change of scale.}
\label{fig:22}
\end{figure}

\begin{figure}
\includegraphics[width=18cm]{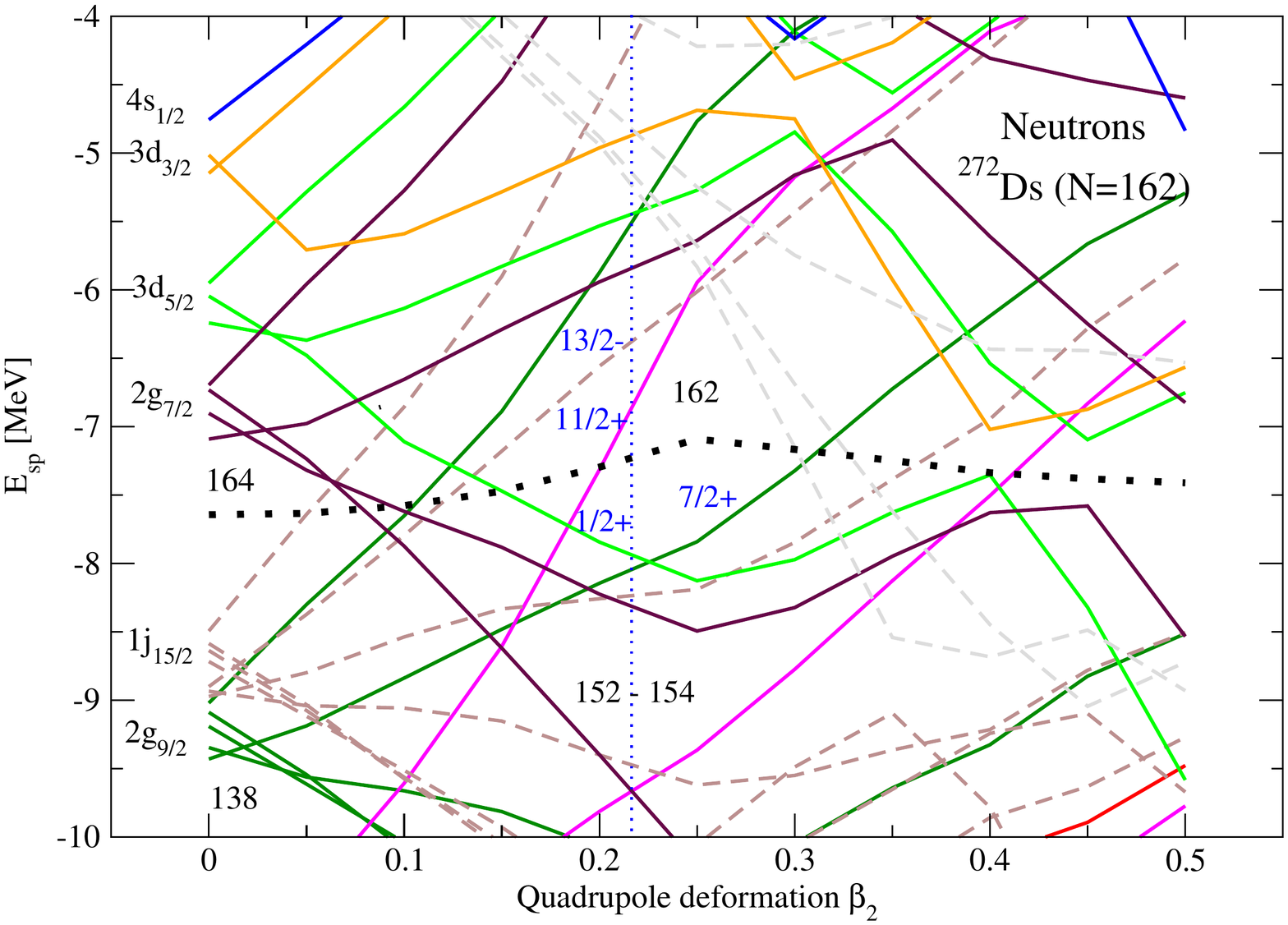}
\centering{}\caption{The same as Fig.~\ref{fig:16} but for $^{\rm 272}$Ds. Note that the y-axis was displaced by 1 MeV without a change of scale.}
\label{fig:23}
\end{figure}

The portion of the single-particle spectra shown in Figs.~\ref{fig:16} --\ref{fig:23} includes  four major neutron orbitals present in the region, the high spin, positive parity  $2g_{9/2}$ and the negative parity $1j_{15/2}$, and low spin, positive parity $3d_{5/2}$ and $2g_{7/2}$. Components of spherical multiplets, $1i_{11/2}$ (just above N=126) and  $2h_{11/2}$ (first above N=184), reaching this region at higher deformation are also shown. We find the spherical N=126 and N=184 shell gaps are well reproduced and,  in addition, shell gaps at  N=138 and N=164 are predicted. 

The sequence of spherical shells $1i_{11/2}$, $2g_{9/2}$, $1j_{15/2}$, $2g_{7/2}$ and $3d_{5/2}$ predicted in the QMC model is not exactly the same as   traditionally used \cite{nilsson1969,chasman1977}. The $2g_{9/2}$ orbital is shifted above the $1i_{11/2}$ by about 2 MeV and the $3d_{5/2}$  is above $2g_{7/2}$ by about 0.7 MeV. On the other hand, the QMC prediction agrees, for example, with the ordering calculated with the SLy4 Skyrme interaction and the relativistic mean field calculation with NL1 force of $^{252}$No \cite{afanasjev2010}. Dobaczewski et al. \cite{dobaczewski2015} illustrated, using an example of $^{254}$No and a variety of mean-field models, that the level ordering at sphericity is model dependent and it affects opening and closing shell gaps in the deformed region. Asai et al., (\cite{asai2015} and references therein) noted that that the recoil term, included in mean-field models which is not present in most macro–-micro calculations, does have non-negligible effects on single-particle energies.
 
The calculated Fermi level is indicated by a horizontal thick dotted (black) line and the ground state deformation by a vertical thin dotted (blue)  line for each isotope. At spherical shapes, the orbitals are labeled with their spherical quantum numbers. At deformation, orbitals are identified by their spherical origin. Close to the Fermi surface they are labeled by $\Omega^\pi$ and the number of neutrons in a simple filling approximation. The sharp changes in the slope of the single-particle energies at high deformation do not affect the QMC results and have not been analysed in detail.

More detailed investigation of validity of the neutron single-particle spectra, as predicted by QMC, requires a comparison of experimental data with the QMC spin-parity assignments. Such data can be obtained from ground (or isomeric) spin-parity assignments in adjacent odd-A nuclei. As the present calculation provides single-particle energies only in even-even nuclei which, as already mentioned in Sec.\ref{sps}, such comparison is inevitably only qualitative. With this caveat, we present in Table~\ref{tab:4} single-particle  energies normalized to the Fermi level within $\pm$ 1 MeV (1.7 MeV for $^{268}$Hs). The $\Omega^\pi$ values are compared for FRDM results \cite{moller2019} and  experimental spin-parity assignments of N$\pm$1 nuclei \cite{nndc} which are however assigned in a model dependent way. The table indicates that some of the orbitals, predicted by QMC close to the Fermi surface, follow a general experimental trend. However, more conclusive statement has to be deferred to future extension of the QMC model to odd-A nuclei.

Examination of the single-particle spectra in Figs.~\ref{fig:16} --\ref{fig:23} suggests a (model dependent) development of the energy gaps around N=152 and N=162 starting $^{252}$Fm. The N=152 weakens with increasing Z and N in the contrary with the N=162 gap, most prominent $^{268}$Hs and $^{272}$Ds. There is no clear evidence of the energy gaps in $^{244}$Cm and $^{248}$Cf.

\section{Summary and outlook}
\label{summary}

The underlying concept of this work has been to explore the sensitivity of the non-relativistic mean-field models, represented by the  QMC as an example, to the shell structure of even-even (semi)magic nuclei below and including $^{208}$Pb as well as SHE nuclei in the region  96 $<$ Z $<$110. The results have been compared to macro-microscopic model.

The effective QMC EDF has a different density dependence than the frequently used Skyrme or Gogny EDF \cite{guichon2018,stone2007} but is used in the same Hartree-Fock + BCS approximation technique.  As compared to the more traditional EDF, the QMC EDF has fewer parameters which are well constrained within physically justified limits\cite{guichon2018}.

As discussed in Secs.~\ref{calmeth}, the HF+BCS model, starting from a set of trial wavefunctions, provides self-consistently single-particle states, their energies and occupation probabilities. The resulting single-particle potential represents the actual density distribution in a given nucleus on the mean field  level, but does not include  important short-range correlations existing in real nuclei. The single-particle wavefunctions are used to calculate observables of the nuclear ground state such as binding energies, density distributions, two-neutron-separation energies, $Q_\alpha$ values etc. The key point is that the wavefunctions of {\em all occupied states} and involved in determination of the bulk ground state properties. It follows that inevitably the sensitivity to small variations in individual single-particle states is reduced. This effect increases with increasing number of nucleons in a nucleus when the density of states grows. Therefore, one has to investigate the single-particle states themselves to seek for changes in shell structure.  We believe this is a generic feature of mean-field models.

Indeed, we found in this work  that bulk properties based on ground state binding energies has not yield a signal of the N=152 and 162 deformed shell gaps. On the other hand,  the investigation of the single-particle spectra and, to some extent, pairing energies, revealed a suggestive evidence for the shell gaps around N=152 and N=162, as predicted by macro-micro models and, so far, confirmed in cases accessible to experiment.

In the macro-micro models the total potential energy is a sum of the macroscopic term and the microscopic term, representing the shell-plus-pairing correction, calculated as a function of shape and proton and neutron number (for more details see \cite{moller2016}). The single-particle potential felt by a nucleon is calculated as a sum of the spin-independent nuclear part, calculated in terms of the folded-Yukawa potential, the spin-orbit potential, and a Coulomb potential. The basis functions used to generate the matrix elements of the single-particle Hamiltonian is a set of deformed, axially symmetric, harmonic-oscillator eigenfunctions. In the most recent calculation, the Lipkin-Nogami model is being used to calculate the pairing effects and the Strutinsky method to calculate shell effects. There are 38(36) constants in the FRDM(FRLDM) models in the expressions in the models, adjusted to nuclear masses, mass-like quantities and other considerations \cite{moller2016}. The impressive predicted power of the models has been tested not only on nuclear masses, but also on other observables, such as ground state spins of odd-A and odd-odd nuclei, beta decay properties and fission barriers \cite{moller2012,moller2016,moller2019}).  It will be interesting to compare its prediction to experiment in the SHE region when more data will become available.

The obvious future  development of the application of the QMC model in the region of SHE is an extension to odd-A  and odd-odd nuclei, which is currently under way. This will allow us to build a more comprehensive picture of the region studied in this work and give confidence in a further search for islands of stability. It will also provide a further testing ground fo the predictive power of the model. We do not expect to achieve as impressive agreement with experiment as the macro-micro models using a model with 4+2 adjustable parameters as compared to tens for constants and empirical relations in macro-micro models although we are encouraged with the results so far.  However, the new physics in the QMC model may provide an inspiration for future development of low-energy nuclear structure models.

\section{Acknowledgement}
J.R.S. gratefully acknowledges Emiko Hyama's invitation to the Kyushu University where this project has been initiated, and the hospitality and financial support during the visit. This work was also supported in part by the University of Adelaide and by the Australian Research Council through the ARC Centre of Excellence for Particle Physics at the Terascale (CE110001104) and Discovery Projects DP150103101 and DP180100497.


\clearpage
\vskip 2 cm
\centering
\begin{table}
\caption{\label{tab:1} Differences between the experimental and calculated ground state binding energies in MeV as calculated  in QMC, FDRM \cite{moller2016} and MM \cite{muntian2003,muntian2003a} models. Experimental data taken from \cite{wang2017}.}
\vspace{4mm}
\begin{tabular}{p{1.5cm}p{1.5cm}p{1.5cm}p{1.5cm}p{1.5cm}|p{0.8cm}p{1.5cm}p{1.5cm}p{1.5cm}p{1.5cm}p{1.5cm}} \hline
      Z    &    N    &     QMC    &    FRDM   &     MM   &  &Z    &    N    &     QMC    &    FRDM   &     MM   \\ \hline
    98    &   146  &     -1.63   &    -0.09   &    $-$    & &102   &   150 &   -1.57   &   -0.69  &   0.08   \\  
           &    148  &     -1.98   &    -0.16  &    $-$    &  &       &    152 &    -1.61   &     -0.64  &   -0.02 \\
           &    150  &     -2.19   &    -0.29  &    $-$    &   &     &      154 &    -2.20   &     -0.55  &   -0.06 \\
           &    152  &     -2.47   &    -0.51 &    $-$    &   &104 &       152 &    -1.34  &     -0.65  &    0.08 \\  
           &    154  &     -3.17   &    -0.42  &    $-$    &   &     &       154  &    -1.83  &    -0.55  &    0.07\\
           &    156  &     -3.72   &    -0.31  &    $-$    &  &106 &       154  &   -1.08   &    -0.66  &   0.13 \\
  100   &     146  &     -0.90   &    -0.48  &    $-$    &   &      &       156   &   -1.46   &    -0.60 &   -0.04 \\
          &     148	 &    -1.30    &    -0.51   &    $-$    &  & 108 &      156   &	-0.79   &    -0.75 &   0.13  \\
          &     150	 &    -1.52    &    -0.57  &    $-$    &   &       &       158  &	-1.18   &   -1.18  &  -0.34 \\
          &     152	&     -1.56   &     -0.50  &    $-$    &   &110&        160  &    -1.13  &     -1.88  &  -0.09 \\
          &     154 &     -2.35   &      -0.52 &    $-$    &    &    &                &              &                &            \\
          &     156	&      0.30   &      -0.44  &    $-$    &    &    &                &              &                &        \\ \hline
\end{tabular}
\end{table}

\clearpage
\vskip 2 cm
\centering
\begin{longtable}{*4{p{3.5cm}}}
\caption{\label{tab:2} Calculated and experimental proton single-particle energies. Experimental data were taken from \cite{grawe2007} except for $^{\rm 90}$Zr \cite{bespalova2005}. Complementary data can be found in \cite{grawe2001,schwierz2007,mahaux1991,cao2014}. The entries marked with an asterisk indicate that the wave functions contains two components as explained in Sec.~\ref{calmeth}}\\

\hline \multicolumn{1}{l}{Nucleus} & \multicolumn{1}{l}{configuration} & \multicolumn{1}{l}{ E$^{cal}_{sp}$ [MeV]}&  \multicolumn{1}{l}{E$^{exp}_{sp}$} \\ \hline 
\endfirsthead

\multicolumn{4}{r}%
{{ \tablename\ \thetable{} -- continued from previous page}} \\

\hline \multicolumn{1}{l}{Nucleus} & \multicolumn{1}{l}{configuration} & \multicolumn{1}{l}{ E$^{cal}_{sp}$ [MeV]}&  \multicolumn{1}{l}{E$^{exp}_{sp}$} \\ \hline 
\endhead

\hline \multicolumn{4}{r}{{Continued on next page}} \\ \hline
\endfoot

\hline
\endlastfoot

$^{\rm 40}$Ca & 1d$_{5/2} $& -13.57   &  -13.73  \\
                 &  1d$_{3/2} $    & -8.47\text{*} & -8.33   \\
                &  2s$_{1/2} $     & -8.27  & -10.85 \\
                &  1f$_{7/2} $     &-1.46  & -1.09 \\  \hline
  $^{\rm 48}$Ca & 1d$_{5/2} $& -21.019   &  -21.58    \\
                &  1d$_{3/2} $    & -16.273\text{*} & -16.17      \\
              &  2s$_{1/2} $     & -14.95  & -15.81    \\
              &  1f$_{7/2} $     & -8.78 & -9.63    \\
                 &  1f$_{5/2} $  & -2.43\text{*}   &  -4.55    \\
                 &  2p$_{3/2} $  & -2.20   &  -6.55    \\
               &  2p$_{1/2} $  & -1.00\text{*}   &  -5.05    \\  \hline
 $^{\rm 56}$Ni  &  1d$_{3/2} $    & -13.45\text{*} & -10.080      \\
              &  2s$_{1/2} $     & -12.30  & -10.72    \\
               &  1f$_{7/2} $     & -5.81 & -7.160    \\
                 &  2p$_{3/2} $     & -0.63  & -0.740    \\ \hline
  $^{\rm 78}$Ni&   1f$_{7/2} $     & -19.43& -20.06   \\
                &  1f$_{5/2} $    & -14.60\text{*} & -14.94      \\
               &  2p$_{3/2} $     & -13.52  & -13.44    \\
                &  2p$_{1/2} $  & -12.26\text{*} &  -12.04   \\
              &  1g$_{9/2} $     & -8.27  & -8.91    \\   \hline
$^{\rm 90}$Zr&   1f$_{7/2} $     & -15.03& -15.56(155) \\
                &  1f$_{5/2} $    & -10.53\text{*} & -10.37(110)  \\
               &  2p$_{3/2} $     & -8.35  & -10.11(110) \\
                &  2p$_{1/2} $  & -6.99\text{*} &  -6.97(70) \\
              &  1g$_{9/2} $     & -4.86& -5.41(54) \\ \hline
 $^{\rm 100}$Sn  &  1f$_{5/2} $    & -7.78\text{*} & -8.71   \\
               &  2p$_{3/2} $     & -5.68 & -6.38 \\
                &  2p$_{1/2} $  & -4.40\text{*} &  -3.53  \\
               &  1g$_{9/2} $     & -2.00  & -2.92   \\  \hline
 $^{132}$Sn   &  2p$_{1/2} $  & -16.26\text{*} &  -16.13   \\
               &  1g$_{9/2} $     & -14.19  & -15.78    \\ 
           &  1g$_{7/2} $     & -9.53\text{*} & -9.65    \\ 
             &  2d$_{5/2} $     & -6.549  & -8.69    \\  
            &  2d$_{3/2} $     & -5.56\text{*}  & -6.95    \\ 
          &  1h$_{11/2} $     & -4.62  & -6.86\text{*}    \\  \hline
  $^{\rm 208}$Pb  &  1g$_{7/2} $  & -13.23 & -11.49    \\ 
           &  2d$_{5/2} $     & -9.58  & -9.70    \\ 
         &  1h$_{11/2} $     & -8.41  & -9.36    \\ 
           &  2d$_{3/2} $     & -8.17\text{*}  & -8.36    \\ 
          &  3s$_{1/2} $     & -6.84& -8.01    \\ 
         &  1h$_{9/2} $     & -4.04\text{*}& -3.80    \\ 
       &  2f$_{7/2} $     & -0.90& -2.90   \\ 
         &  1i$_{13/2} $     & -0.13& -2.19    \\  \hline
  \end{longtable}

\clearpage
\vskip 2 cm
\centering
\begin{longtable}{*4{p{3.0cm}}}
\caption{\label{tab:3} The same as Table~\ref{tab:2} but for neutron single-particle states.}  \\

\hline \multicolumn{1}{l}{Nucleus} & \multicolumn{1}{l}{configuration} & \multicolumn{1}{l}{ E$^{cal}_{sp}$ [MeV]}&  \multicolumn{1}{l}{E$^{exp}_{sp}$} \\ \hline 
\endfirsthead

\multicolumn{4}{r}%
{{ \tablename\ \thetable{} -- continued from previous page}} \\

\hline \multicolumn{1}{l}{Nucleus} & \multicolumn{1}{l}{configuration} & \multicolumn{1}{l}{ E$^{cal}_{sp}$ [MeV]}&  \multicolumn{1}{l}{E$^{exp}_{sp}$} \\ \hline 
\endhead

\hline \multicolumn{4}{r}{{Continued on next page}} \\ \hline
\endfoot

\hline
\endlastfoot

$^{\rm 40}$Ca & 1d$_{5/2} $& -21.31   &  -21.27 \\
                 &  1d$_{3/2} $    & -16.03\text{*} & -15.64      \\
                &  2s$_{1/2} $     & -15.99  & -18.11    \\
                &  1f$_{7/2} $     &-8.84 & -8.36   \\
                &  2p$_{3/2} $     & -3.79 & -6.42    \\
                 &  2p$_{1/2} $     & -2.26\text{*}  & -4.42    \\
                  &  1f$_{5/2} $     & -1.87\text{*} & -2.65    \\ \hline
 $^{\rm 48}$Ca &  1f$_{7/2} $     & -9.32 & -9.95    \\
                 &  2p$_{3/2} $     & -4.85  & -5.15    \\
                 &  2p$_{1/2} $  & -3.37\text{*}  &  -3.12    \\
                 &  1f$_{5/2} $  & -2.30\text{*}   &  -1.20    \\ \hline
 $^{\rm 56}$Ni  &  1d$_{3/2} $    & -23.43\text{*} & -19.84   \\
              &  2s$_{1/2} $     & -22.44 & -20.40    \\
               &  1f$_{7/2} $     & -15.57 & -16.65   \\
                 &  2p$_{3/2} $     & -10.25  & -10.25    \\
                 &  1f$_{5/2} $  & -9.06\text{*}   &  -9.48    \\
                 &  2p$_{1/2} $  & -8.81\text{*} &  -9.14    \\ 
           &  1g$_{9/2} $  & -3.59 &  -6.55    \\ \hline
 $^{78}$Ni  &  1f$_{5/2} $    & -10.83\text{*} & -8.39     \\
               &  2p$_{3/2} $     & -11.00  & -8.54   \\
                &  2p$_{1/2} $  & -9.21\text{*} &  -7.21  \\
              &  1g$_{9/2} $     & -5.66 & -5.86  \\
             &  2d$_{5/2} $     & -0.89 & -2.21   \\   \hline
$^{\rm 90}$Zr&   1f$_{7/2} $     & -22.17 & -21.27(213)\\
                &  1f$_{5/2} $    & -17.61\text{*}& -14.48(145)  \\
               &  2p$_{3/2} $     & -16.59 & -13.85(138) \\
                &  2p$_{1/2} $  & -15.01\text{*}&  -13.19(132) \\
              &  1g$_{9/2} $     & -11.93  & -12.15(120) \\ 
              &  2d$_{5/2} $     & -6.07& -6.85(70) \\ 
             &  3s$_{1/2} $     & -4.11  &  -5.63(56)\\ 
             &  2d$_{3/2} $     & -3.97\text{*}  & 4.70(47)\\  \hline
$^{\rm 100}$Sn &  2p$_{1/2} $  & -19.39\text{*} &  -18.38   \\
               &  1g$_{9/2} $     & -16.56  & -17.93    \\ 
              &  2d$_{5/2} $     & -10.14  & -11.15    \\ 
            &  1g$_{7/2} $     & -10.41\text{*}  & -11.07    \\ 
              & 3s$_{1/2} $     & -7.75  & -9.60    \\ 
            &  2d$_{3/2} $     & -8.12\text{*}  & -9.50    \\ 
             &  1h$_{11/2} $     & -6.13  & -8.60    \\ \hline
   $^{\rm 132}$Sn  &  1g$_{7/2} $     & -12.187 & -10.28    \\ 
           &  2d$_{5/2} $     & -10.74 & -9.04    \\ 
         &  3s$_{1/2} $     & -8.44  & -7.72   \\ 
        &  1h$_{11/2} $     & -7.70  & -7.46   \\ 
         &  2d$_{3/2} $     & -8.67\text{*}  & -7.39    \\ 
          &  2f$_{7/2} $     & -1.610  & -2.45   \\ 
         &  3p$_{3/2} $     & -0.115  & -1.59   \\ 
        &  1h$_{9/2} $     & -0.80\text{*} & -0.88    \\ 
         &  3p$_{1/2} $     & -0.77\text{*}& -0.79  \\ \hline
 $^{\rm 208}$Pb   &  1h$_{9/2}$     & -13.24\text{*}& -10.78    \\ 
       &  2f$_{7/2} $     & -11.27& -9.71    \\ 
        &  1i$_{13/2} $     & -9.80& -9.00    \\ 
       &  2f$_{5/2} $     & -8.92\text{*}& -7.94    \\ 
       &  3p$_{3/2} $     & -8.34& -8.27    \\ 
       &  3p$_{1/2} $     & -7.47\text{*}& -7.37    \\   
         &  1i$_{11/2} $     & -3.52\text{*}& -3.16    \\ 
       &  2g$_{9/2} $     & -2.91& -3.94   \\ 
       &  1j$_{15/2} $     & -1.37& -2.51    \\ 
      &  2g$_{7/2} $     & -0.26\text{*}& -1.40    \\ 
       &  3d$_{5/2} $     & -0.41& -2.37    \\ 
       &  4s$_{1/2} $     & -0.045& -1.90    \\ \hline  
 
\end{longtable}

\clearpage
\vskip 2 cm
\centering
\begin{table}
\caption{\label{tab:4}Neutron single-particle orbitals in the vicinity of the Fermi surface in $^{\rm 244}$Cm, $^{\rm 248}$Cf, $^{\rm 252}$Fm, $^{\rm 256}$No, $^{\rm 260}$Rf, $^{\rm 264}$Sg, $^{\rm 268}$Hs and $^{\rm 272}$Ds calculated with respect to the Fermi energy. Experimental data on spin-parity assignment to ground states odd-A nuclei with N$\pm$1 (where available) \cite{nndc} and results of the  FRDM calculation are included for comparison. For more information see text and figures \ref{fig:16} --\ref{fig:23}.}
\vspace{4mm}
\scalebox{0.9}{
\begin{tabular}{cccccccc} \hline
Z    &    N   &   Energy [MeV] & Spherical &   $\Omega^\pi$   &   N$_{\rm odd}$ &  $\Omega^\pi$ (FDRM)  &  exp  \\ \hline  
 96   &   148  &  -0.589        &    2g$_{7/2} $  &    1/2$^+$     &     147            &   5/2$^+$  & 5/2$^+$  \\
        &           &  -0.482        &    1j$_{15/2} $  &    7/2$^-$     &     149            &   7/2$^+$  & 7/2$^+$  \\
        &           &  0.001          &    2g$_{9/2} $  &    5/2$^+$    &                        &            &           \\
        &           &  0.876          &    1j$_{15/2} $  &    9/2$^-$    &                        &            &           \\  \hline
 98   &   150  &  -0.919        &    1j$_{15/2} $  &   7/2$^-$    &      149             &  7/2$^+$  & 9/2$^-$  \\
        &           &     -0.427     &    2g$_{9/2} $   &   5/2$^+$    &      151            &   9/2$^-$   & 9/2$^-$  \\
         &          &        0.379    &    1j$_{15/2} $  &   9/2$^-$    &                        &            &           \\
         &          &        0.403    &    1i$_{11/2} $  &   9/2$^+$   &                        &            &           \\    \hline
100  &    152  &    -0.881     &    2g$_{9/2} $   &  5/2$^+$    &      151            & 9/2$^-$ & (9/2$^-$) \\
        &            &    -0.126     &    1j$_{15/2} $  &   9/2$^-$    &      153            & 7/2$^+$ & 1/2$^+$  \\
        &            &    -0.022     &    1i$_{11/2} $  &   9/2$^+$   &                        &            &           \\
        &            &    1.006       &    2g$_{7/2} $  &   3/2$^+$   &                        &            &           \\     \hline
102  &    154  &    -0.805     &    1j$_{15/2} $  &   7/2$^-$   &       153            & 7/2$^+$ & ( 1/2$^+$)  \\
        &            &    -0.430     &    1i$_{11/2} $  &   9/2$^+$  &        155            &11/2$^-$ &( 7/2$^+$)   \\
        &            &    0.519       &    2g$_{7/2} $  &   3/2$^+$  &                        &            &           \\
        &            &    0.710       &    3d$_{5/2} $  &    1/2$^+$ &                        &            &           \\    \hline
104  &    156  &   -0.821      &    1i$_{11/2} $  &   9/2$^+$  &       155            & 7/2$^+$&           \\
        &            &    0.052      &     2g$_{7/2} $  &    3/2$^+$ &        157            &9/2$^+$&           \\
        &            &   0.287       &    3d$_{5/2} $   &    1/2$^+$ &                        &            &           \\
        &            &   0.371       &    1j$_{15/2} $  &    11/2$^-$ &                        &            &           \\   \hline
106  &    158  &   -0.388     &    2g$_{7/2} $  &    3/2$^+$   &      157            &9/2$^+$ &           \\
        &            &   -0.106     &    3d$_{5/2} $  &    1/2$^+$   &      159            &1/2$^+$ &           \\
        &            &   -0.074     &    1j$_{15/2} $ &    11/2$^-$  &                        &            &           \\
        &            &    0.150      &    2g$_{9/2} $ &    7/2$^+$   &                        &            &           \\   \hline
108  &    160  &   -0.865    &    2g$_{7/2} $  &    3/2$^+$    &     159            & 1/2$^+$ & (3/2$^+$) \\
        &            &   -0.538    &    3d$_{5/2} $   &    1/2$^+$   &     161            & 3/2$^+$ &                  \\
        &            &   -0.551    &     1j$_{15/2} $   &   11/2$^-$ &                        &            &           \\
        &            &   1.532      &    1i$_{11/2} $   &    11/2$^+$ &                        &            &           \\
        &            &   1.610      &   1j$_{15/2} $    &    13/2$^-$ &                        &            &           \\  \hline
110  &    162  &   -0.849    &    2g$_{9/2} $   &    7/2$^+$   &    161            &  3/2$^+$ &           \\
       &            &   -0.548     &    3d$_{5/2} $   &    1/2$^+$   &    163            & 13/2$^-$ &           \\
       &            &   -0.019    &     1i$_{11/2} $   &    11/2$^+$ &                        &            &           \\
       &            &  0.738       &     1j$_{15/2} $  &     13/2$^-$ &                        &            &           \\   \hline

\end{tabular}
}
\end{table}

\clearpage

\clearpage

\clearpage

\clearpage

\clearpage

\clearpage

\clearpage

\clearpage

\clearpage

\clearpage

\clearpage

\clearpage

\clearpage

\clearpage

\clearpage

\clearpage

\clearpage

\clearpage

\clearpage

\clearpage

\clearpage

\clearpage

\clearpage

\end{document}